\documentclass[a4paper,11pt]{article}

\pdfoutput=1 % if your are submitting a pdflatex (i.e. if you have images in pdf, png or jpg format)
\usepackage{jheppub} % for details on the use of the package, please see the JHEP-author-manual

\author[a,b]{Y. Reyimuaji}
\author[a,b]{and A. Romanino}

\affiliation[a]{\SISSA}
\affiliation[b]{\ICTP}

\emailAdd{yreyimua@sissa.it}

\newcommand{\rep}{U}
\newcommand{\U}{U_{\text{PMNS}}}
\newcommand{\V}{V_{\text{CKM}}}

\newcommand{\Tab}[1]{Table~\ref{tab:#1}}
\newcommand{\tab}[1]{table~\ref{tab:#1}}
\newlength{\myl}
\DeclareMathOperator{\bdiag}{BDiag}

\newcommand{\titletext}{Can an unbroken flavour symmetry provide an approximate description of lepton masses and mixing?} 

\newcommand{\abstracttext}{We provide a complete answer to the following question: what are the flavour groups and representations providing, in the symmetric limit, an approximate description of lepton masses and mixings? We assume that neutrinos masses are described by the Weinberg operator. We show that the pattern of lepton masses and mixings only depends on the dimension, type (real, pseudoreal, complex), and equivalence of the irreducible components of the flavour representation, and we find only six viable cases. In all cases the neutrinos are either anarchical or have an inverted hierarchical spectrum. In the context of SU(5) unification, only the anarchical option is allowed. Therefore, if the hint of a normal hierarchical spectrum were confirmed, we would conclude (under the above assumption) that symmetry breaking effects must play a primary role in the understanding of neutrino flavour observables. In order to obtain the above results, we develop a simple algorithm to determine the form of the lepton masses and mixings directly from the structure of the decomposition of the flavour representation in irreducible components, without the need to specify the form of the lepton mass matrices.}

\newcommand{\SISSA}{SISSA/ISAS and INFN, I--34136 Trieste, Italy}
\newcommand{\ICTP}{ICTP, Strada Costiera 11, I--34151 Trieste, Italy}

\newcommand{\ord}[1]{\mathcal{O}\left( #1 \right)}

\newcommand{\dm}[1]{{\Delta m^2_{#1}}}

\newcommand{\Eq}[1]{Eq.~(\ref{eq:#1})}
\newcommand{\eq}[1]{eq.~(\ref{eq:#1})}
\newcommand{\Eqs}[1]{Eqs.~(\ref{eq:#1})}

\DeclareMathOperator{\diag}{Diag}

\newlength{\myem}
\title{\boldmath \titletext}
\abstract{\abstracttext}

\begin{document}

\maketitle
\flushbottom

\section{Introduction}
\label{sec:intro}

One of the most popular attempts at understanding the Standard Model (SM) fermion mass and mixing pattern makes use of ``flavour'' symmetry groups~\cite{Froggatt:1978nt,Bijnens:1986tt,Leurer:1992wg,Dine:1993np,Ibanez:1994ig,Pomarol:1995xc,Barbieri:1995uv,Carone:1995xw,Dudas:1996fe,Barbieri:1996ww,Barbieri:1997tu,Carone:1997qg,Irges:1998ax,Elwood:1998kf,Ferretti:2006df}. The flavour symmetry $G$ is spontaneously broken to a subgroup $H$ (trivial if $G$ is completely broken) and the source of breaking is provided by the vacuum expectation value (vev) of one or more scalar fields (``flavons''), singlets under the SM, but transforming non-trivially under $G$. Focussing on lepton flavour, and denoting by $m_E$ and $m_\nu$ the charged lepton and neutrino mass matrices, we then have 
\begin{equation}
\begin{aligned}
m_E &= m^{(0)}_E + m^{(1)}_E \\
m_\nu &= m^{(0)}_\nu + m^{(1)}_\nu 
\end{aligned} \;,
\label{eq:main}
\end{equation}
where $m^{(0)}_E$, $m^{(0)}_\nu$ are invariant under $G$ and therefore survive in the limit in which the flavour symmetry is not broken, while $m^{(1)}_E$, $m^{(1)}_\nu$ are generated by the vevs of the flavons, are invariant under $H$ but not under $G$, and vanish in the symmetric limit. The non-vanishing entries in $m^{(0)}_E$, $m^{(0)}_\nu$ are often, and here, assumed to be of the same order, according to the principle that flavour hierarchies should be accounted for by the flavour model itself. The size of the corrections associated to the symmetry breaking effects is assumed to be smaller (except of course when the leading order term vanishes). 

Within the above scheme, we are interested in assessing whether the leading order pattern of lepton masses and mixings is accounted for by the flavour symmetry itself or rather by the symmetry breaking effects.  We can indeed distinguish two distinct cases. 
\begin{enumerate}
\item
The symmetric form of the mass matrices, $m^{(0)}_E$ and $m^{(0)}_\nu$, provides an approximate description of lepton flavour observables, in particular of the PMNS matrix; $m^{(1)}_E$ and $m^{(1)}_\nu$ 
provide the moderate correction necessary for an accurate description. In such a case, we can say that the leading order pattern of lepton masses and mixings is accounted for by the flavour symmetry itself. 
\item
The symmetry breaking corrections are important even for an approximately correct description of lepton flavour observables. If the size of corrections is assumed to be smaller than the non-vanishing symmetric terms, this can happen if $m^{(0)}_E$ or $m^{(0)}_\nu$ vanishes, in which case the PMNS matrix is fully undetermined in the symmetric limit; or in the presence of an accidental enhancement of the role of $m^{(1)}_E$, $m^{(1)}_\nu$.\footnote{This is the case for example if one of the neutrino masses obtained in the symmetric limit is accidentally suppressed and ends up being of the same order of the smaller symmetry breaking corrections. In such a case, the symmetric limit prediction for some of the lepton mixing angles can be drastically modified, and actually determined, by the symmetry breaking effects~\cite{Binetruy:1996xk,Binetruy:1996cs,Grossman:1998jj,Altarelli:2002hx}.}
\end{enumerate}

In this paper, we aim at providing a complete study of the first case, thus also assessing the need to resort to the second possibility. We will namely obtain a complete characterisation of the flavour symmetry groups $G$ (of any type) and their representations on the SM leptons providing an approximate understanding of lepton masses and mixing in the symmetric limit. Moreover, we will show that the results can be extended to the second case as well, if some (non-trivial) hypotheses hold. 

The first case has been extensively considered since the earliest attempts of understanding the pattern of fermion masses and mixings. As charged fermion masses show a clear hierarchical structure, it is natural to account for the lightness of the first two families in terms of small symmetry breaking effects. For example, the symmetric limit could allow the third family to acquire a mass but not the first two. The symmetric limit is then close to what observed, with the small Yukawas associated to the lighter families approximated by zero and the CKM matrix approximated by the identity matrix. The lighter masses and the small CKM mixings are then generated by small perturbations of the symmetric limit associated to the spontaneous breaking of the flavour symmetry. 

Does the above scheme apply to neutrino masses and  mixings as well? While many models have been proposed in which it does, to our knowledge a systematic analysis charting all possibilities is missing. Given the large variety of possible cases, it is not a priori obvious that a complete analysis can be carried out in an effective way and would produce results that can be expressed in a concise form. Interestingly, this turns out to be the case: the problem can be studied in full generality, admits a precise mathematical formulation, and a complete and compact solution. While specific implementations of the full solution are well known, the analysis shows that the options we will find are the only possible ones, thus providing a final answer to the above question. The mathematical formulation of the problem, and the definition of ``approximate description'' will be discussed in section~\ref{sec:patterns}. 

As we will see, while the possibility that lepton flavour can be approximately understood in terms of a symmetry principle alone is aesthetically appealing, future data might disfavour it. In such a case, the symmetry breaking effects become essential for an understanding of lepton flavour. One can then wonder whether the knowledge of the symmetry breaking pattern $G\to H$ can be sufficient, or the intricacies of the flavon spectrum, vevs, and potential should be specified. The knowledge of the breaking patter is sufficient if $m^{(0)}_E$ or $m^{(0)}_\nu$ vanishes in the symmetric limit and the corrections $m^{(1)}_E$, $m^{(1)}_\nu$ are in the most general form allowed by the residual symmetry $H$, with all their entries of the same order. Under such a (non-trivial) hypothesis, it turns out that the techniques developed to study the symmetric limit can be easily extended to study this case as well, and that the conclusions do not change. 

The analysis we perform is fully general in the assumptions that i) the light neutrino masses are in Majorana form and ii) the symmetry arguments can be applied directly to the light neutrino mass matrix (or to the Weinberg operator from which it originates). The second assumption is relevant in the case in which the light neutrino mass matrix arises from physics well above the electroweak scale, the prototypical case being the integration of heavy singlet neutrinos in the context of the seesaw mechanism. In such a case, the heavy degrees of freedom also transform under the flavour symmetry, and a symmetric limit can be defined for their mass matrix as well. One can then wonder whether the ``low energy'' analysis performed in terms of the light neutrino mass matrix captures the features of the full analysis. This turns out to be often true, but a necessary condition is that the mass matrix of heavy neutrinos be non-singular in the symmetric limit. A thorough study of such important caveat will be the subject of a separate work.  

The paper is structured as follows. Section~\ref{sec:patterns} contains the main result of this paper, i.e.\ the classification of flavour groups and representations leading to an approximate description of lepton masses and mixings. Section~\ref{sec:breaking} discusses the case in which either the neutrino or the charged lepton masses all vanish in the symmetric limit, and lepton mixing is determined by symmetry breaking effects. Section~\ref{sec:unification} investigate the additional constraints provided by grand-unification. In section~\ref{sec:conclusions} we draw our conclusions.

\section{Lepton masses and mixings in the symmetric limit}
\label{sec:patterns}

In this section, we aim at providing a full characterisation of the flavour groups $G$ and their representations on the leptons leading to a viable approximate description of lepton masses and mixings in the symmetric limit. We will proceed in two steps. First, in section~\ref{sec:masses}, we will list all representations leading to an approximate description of lepton masses (but not necessarily of lepton mixing). Then, in section~\ref{sec:mixings}, we will select among them the cases in which the PMNS matrix is also approximately realistic in the symmetric limit. 

Before that, we need to define which lepton mass and mixing patterns we consider an approximate description of what observed and to give a precise mathematical formulation of the problem of finding the groups and representations associated to those patterns. 

The full list of charged lepton and neutrino mass patterns that we consider to be close to what observed is in \tab{masses}. Let us illustrate the table by considering a few examples. The case in which the three charged lepton masses are in the form $(A,0,0)$ can be considered to be close to what observed because of the smallness of the electron and muon masses compared to the tau mass. Only a small correction to that pattern is required in order to provide an accurate description of the charged lepton spectrum. On the contrary, a pattern such as $(A,A,0)$, for example, cannot be considered to be close to what observed, as no pair of charged lepton masses are close to be degenerate. The pattern $(A,B,0)$ is in between. It can be considered close to what observed if $A$ and $B$ are allowed to have different sizes, with $B\ll A$, or vice versa. But not if $A$ and $B$ are assumed to be of the same order of magnitude, unless one entry is accidentally suppressed with respect to the other. In the neutrino sector, a pattern in the form $(a,0,0)$ can be considered to be close to what observed, as only a small correction is required to obtain a realistic normal hierarchical spectrum. The pattern $(0,a,a)$ also provides a good approximate description, as a small correction splitting the two degenerate eigenvalues is only required to obtain a realistic inverted hierarchical spectrum. A normal hierarchical spectrum is at present favoured by data~\cite{Simpson:2017qvj,Capozzi:2017ipn,deSalas:2017kay}, but we still retain the inverted spectrum as a viable possibility. 

\begin{table}
\[
\begin{array}{c||c|c||c|}
& \text{\parbox[c][1.1cm][c]{4cm}{\centering non-zero entries \\ of the same order}}
& \text{\parbox[c][1.1cm][c]{4cm}{\centering hierarchy among non-zero entries}}
& \text{\parbox[c][1.1cm][c]{4cm}{\centering (fully undetermined in the symmetric limit)}} \\
\hline\hline
\text{\parbox[c][1.1cm][c]{1.8cm}{\centering charged \\ leptons}} &
(A,0,0) & \begin{matrix} (A,B,0) \\ (A,B,C) \end{matrix} & (0,0,0) \\
\hline\hline
\text{\parbox[c][1.1cm][c]{1.8cm}{\centering neutrinos \\ NH}} &
(a,0,0) & & \\
\hline
\text{\parbox[c][1.1cm][c]{1.8cm}{\centering neutrinos \\ NH or IH}} &
\begin{matrix} (a,a,a) \\ (a,b,b) \end{matrix} & \begin{matrix} (a,b,0) \\ (a,b,c) \end{matrix} & (0,0,0) \\
\hline
\text{\parbox[c][1.1cm][c]{1.8cm}{\centering neutrinos \\ IH}} &
(0,a,a) & & \\
\hline\hline
\end{array}
\]
\caption{Charged lepton and neutrino mass patterns in the symmetric limit.}
\label{tab:masses}
\end{table}

All the entries in \tab{masses} are assumed to be positive or zero. The last column of the table corresponds to the possibility that the mass spectrum is fully determined by symmetry breaking effects. Such cases will be considered in section~\ref{sec:breaking}. Here, we only need to consider the cases in the first two columns. In the first column we list the cases that can be considered as good leading order approximations even when all the non-zero entries are of the same order of magnitude. The cases in the second column, on the contrary, require some degree of hierarchy or degeneracy between the non-zero entries. Such a distinction is more important for charged leptons than neutrinos. The hierarchy among non-zero entries required in the charged lepton cases to account for the hierarchy $m_e \ll m_\mu \ll m_\tau$ is $\ord{20}$ in the $(A,B,0)$ case and $\ord{200}$ in the $(A,B,C)$ case. On the other hand, in the neutrino case only milder hierarchies up to $\ord{5}$ are required to account for $\dm{12}/|\dm{23}|\ll 1$ in the normal hierarchy case\footnote{For inverted hierarchy, a stronger accidental degeneracy is required. For example, in the $(a,b,0)$ case, $a/|b-a| = \ord{50}$ is required.}. Such a mild hierarchy is not too far from what can be considered to be of the same order. Therefore, we will only care about the distinction between first and second column in the case of charged leptons. In the case of neutrinos, we distinguish the cases leading (after taking into account small symmetry breaking corrections) to a normal hierarchy (NH), an inverted hierarchy (IH), or to any of the two depending on the sizes of the non-zero entries. 

A pedantic remark on the patterns in \tab{masses} (which however will play a role in the following)   concerns the fact that the pattern $(a,b,0)$, for example, includes the case in which $b=a$, as well as the case in which $b = 0$. We define a mass pattern to be ``generic'' if all the entries that are allowed to be different from each other and non-zero are indeed different from each other and non-zero. 

As for the PMNS matrix, we will consider it to be close to what observed in the symmetric limit if either i) none of its elements vanishes or ii) only the 13 element vanishes. Indeed, all of the PMNS entries appear to be of order one, with the exception of the 13 element, $|(\U)_{13}|\approx 0.15$. One of the 21 and 31 elements can be as small as about $0.25$ if leptonic CP violation will turn out be small, unlike what the present fits seem to suggest~\cite{Capozzi:2016rtj,Esteban:2016qun,Abe:2017uxa,Capozzi:2017ipn}. All other elements are bound to be larger than 0.45 ($3\sigma$ bounds from~\cite{Esteban:2016qun}). As a consequence of the above definition, we will not consider PMNS matrices corresponding to a single $2\times 2$ transformation in the 12, 23, or 13 block, which would require at least four matrix entries to vanish. In the case of PMNS matrices obtained by the combination of two $2\times 2$ transformations in different blocks, the PMNS matrix contains one vanishing entry, which is located in the 13 entry if the two $2\times 2$ rotations are in the 23 and 12 block (in this order). 

Having specified the mass and mixing patterns that we consider viable in the symmetric limit, we now want to characterise the flavour groups and representations leading to any of those patterns. Let us then give first of all a precise formulation of the problem. 

The flavour symmetry group $G$ acts on the SM leptons $l_i$ and $e^c_i$ through unitary representations $\rep_l$ and $\rep_{e^c}$ respectively. Here $e^c \sim \overline{e_R}$ denotes the conjugated of the right-handed SM leptons (SU(2)$_L$ singlets with hypercharge $Y=1$), and $l = (\nu,e)^T$ denotes the left-handed leptons (SU(2)$_L$ doublets with hypercharge $Y = -1/2$). With this notation, all the fermion fields are left-handed, which will also turn out to be useful when we will discuss grand-unification in section~\ref{sec:unification}. The charged lepton and neutrino mass matrices arise from the Yukawa and Weinberg operators~\cite{Weinberg:1979sa} respectively,
\begin{equation}
\lambda^E_{ij} e^c_i l_j h^*, \qquad
\frac{c_{ij}}{2\Lambda}l_i l_j h h,
\label{eq:weinberg}
\end{equation}
and are given by
\begin{equation}
m_E = \lambda_E v, \qquad
m_\nu = c\, v^2 /\Lambda ,
\label{eq:massconvention}
\end{equation}
where $h$ is the Higgs field, $v = |\langle h \rangle |$, and Lorentz-invariant contractions of fermion indices are understood. Note the convention in which the singlet leptons appear first in the Yukawa interaction. Note also that the action of $G$ is the same on the two components of $l_i$, $\nu_i$ and $e_i$, as it is supposed to commute with the SM gauge transformations. 

The groups $G$ is arbitrary. It can be continuous or discrete, simple or not, abelian or not, or arbitrary combinations of the above. It is supposed to include all the relevant symmetries, including those possibly used to force specific couplings of the flavons. We denote by $\rep_l$ and $\rep_{e^c}$ its representations on the doublet and singlet leptons respectively. From the invariance of the Yukawa and Weinberg operators, one finds that the lepton mass matrices  $m_E$, $m_\nu$ are invariant if they satisfy
\begin{equation}
m_E = \rep_{e^c}^T(g) \, m_E \, \rep^{\phantom{\dagger}}_l(g) \qquad m_\nu = \rep_l^T(g) \, m_\nu \, \rep^{\phantom{T}}_l(g) \qquad \forall g\in G .
\label{eq:invariance}
\end{equation}
A possible non-trivial transformation of $h$ under $G$ can be reabsorbed in $\rep_l$ and $\rep_{e^c}$. 

We can now formulate the problem we want to address as follows. For each of the $3\times 6 = 18$ combinations of charged lepton and neutrino mass patterns in \tab{masses} (excluding the ones in the third column), we want to determine, or characterise, all groups $G$ and representations $\rep_l$, $\rep_{e^c}$ corresponding to those mass patterns and leading to a viable PMNS matrix. We say that the group and its representation ``correspond to'' or ``force'' a given mass pattern if i) the eigenvalues\footnote{Here and in the following we use ``eigenvalues'' to refer to the singular values of $m_E$, $m_\nu$.} of any pair of invariant matrices $m_E$, $m_\nu$ follow that mass pattern, and if ii) there exists at least a pair of invariant matrices $m_E$, $m_\nu$ such that the eigenvalues not only follow that mass pattern, but are also generic (i.e.\ with all entries that are allowed to be different and non-zero being different and non-zero). The second requirement is needed, as otherwise we could end up with groups and representations corresponding to a different, more constrained, pattern. 

Note that it is important to write the invariance condition for $m_E$, as in \eq{invariance}, and not for $m^\dagger_E m^{\phantom{\dagger}}_E$. In the latter case, the important role of $\rep_{e^c}$ would be lost. 

\subsection{Accounting for lepton masses}
\label{sec:masses}

In this section we characterise all the groups and representations that force each of the 18 combinations of charged lepton and neutrino mass patterns in the first two columns of \tab{masses}. It turns out that it is possible to characterise them in terms of their decompositions into irreducible representations (``irreps''), and of the dimensionality, type (complex, real, or pseudoreal), and equivalence of the irreducible components. 

We remind that a representation is called ``complex'' if it is not equivalent to its conjugated representation. A representation that is equivalent to its conjugated is called ``real'' if it can be represented by real matrices and ``pseudoreal'' if it cannot. Pseudoreal representations have even dimension. 

The full list of irrep decompositions corresponding to a given mass pattern is shown in tables~\ref{tab:irreppattern1}, \ref{tab:irreppattern2}. The first table only contains the charged lepton mass pattern that does not require hierarchies among the non-zero entries, $(A,0,0)$, while the second contains the cases in which a hierarchy is necessary, following the classification in \tab{masses}. In the rest of this section we will prove and illustrate the results in tables~\ref{tab:irreppattern1}, \ref{tab:irreppattern2}. 

In order to prove the results in the tables, we note that there is a close connection between the mass patterns and the irrep decompositions, which we now illustrate. Since the extension is straightforward and useful, let us consider the general case of $n$ lepton families. Let us choose a basis in flavour space in which the charged lepton mass matrix is diagonal, $m_E = \diag(m^E_1\ldots m^E_n)$. In the symmetric limit, the mass eigenvalues are assumed to follow one of the patterns in \tab{masses}, which means that a certain number of them are assumed to be zero (possibly none) and that groups of non-zero masses may be assumed to be degenerate. In full generality, the mass eigenvalues (for both the charged leptons and neutrinos) can then be written in the form 
\begin{equation}
(m_1 \ldots m_n) = 
(\, \overset{d_0}{\overbrace{0 \ldots 0}} \quad
\overset{d_1}{\overbrace{a_1 \ldots a_1}} \quad
\ldots \quad
\overset{d_N}{\overbrace{a_N \ldots a_N}} \,),
\label{eq:defpattern}
\end{equation}
corresponding to a group of $d_0$ vanishing masses and $N$ groups of degenerate masses, with multiplicities $d_1\dots d_N$. In the cases in tables~\ref{tab:irreppattern1}, \ref{tab:irreppattern2}, there is at most one group of degenerate eigenvalues in the neutrino sector, with multiplicity 2 or 3. The values of $a_1\ldots a_N$ can happen to vanish or to be equal to each other. This situation is not generic, though. In a generic set of mass eigenvalues, $a_1\ldots a_N$ are non-zero and all different from each other. 

The results in tables~\ref{tab:irreppattern1}, \ref{tab:irreppattern2} are obtained using the following facts. Consider a given mass pattern, in which charged lepton and neutrino masses are both in the form~\eq{defpattern} (with different multiplicities $d^E_0\ldots d^E_{N_E}$, $d^\nu_0\ldots d^\nu_{N_\nu}$). Then:
\begin{itemize}
\item Each subspace in flavour space associated to (zero or non-zero) degenerate charged lepton masses is invariant under both the representations $\rep_l$ and $\rep_{e^c}$. We can then call $\rep^l_0\ldots \rep^l_{N_E}$ and $\rep^{e^c}_0\ldots \rep^{e^c}_{N_E}$ the representations on those subspaces. 
\item The representations corresponding to non-zero charged lepton masses, $\rep^l_1\ldots \rep^l_{N_E}$ and $\rep^{e^c}_1\ldots \rep^{e^c}_{N_E}$, are conjugated to each other and irreducible. 
\item The representations $\rep^l_0$ and $\rep^{e^c}_0$ corresponding to the set of vanishing masses can be reducible. None of the irreps on which $\rep^{e^c}_0$ decomposes is conjugated to any of the irreps on which $\rep^{l}_0$ decomposes. 
\end{itemize}
The neutrino mass pattern gives further constraints on $\rep_l$:
\begin{itemize}
\item Each set of $d^\nu$ degenerate non-vanishing neutrino masses must correspond to either a real irrep $r=\bar r$ of dimension $d^\nu$; or to a pair of conjugated (Dirac) complex irreps $r+\bar r$ of total even dimension $d^\nu$; or to a pair of equivalent pseudoreal irreps $r+r$ with total dimension $d^\nu$ multiple of four (case hence not relevant with three neutrinos). 
\item The remaining irreps in $\rep_l$ must correspond to the vanishing neutrino masses, and therefore their total dimension should be $d^\nu_0$. Moreover, none of them is real, none of the complex ones is conjugated to any other, and none of the pseudoreal ones is equivalent to any other. 
\end{itemize}

To illustrate how the above remarks lead to the results in tables~\ref{tab:irreppattern1}, \ref{tab:irreppattern2}, let us consider a few examples. Let us first consider the mass pattern $(A,B,C)$ for the charged leptons and $(a,b,c)$ for the neutrinos. As we have three different non-vanishing charged lepton masses, $\rep_l$ must decompose into 3 one-dimensional irreps and $\rep_{e^c}$ into the three conjugated ones. As we have three different non-vanishing neutrino masses, the three one-dimensional representations in which $\rep_l$ decomposes must be real. Depending on whether the three real irreps are equivalent or not, we find the three cases listed in table~\ref{tab:irreppattern2}. The last case, corresponding to $\rep_l \sim \rep_{e^c} \sim 1+1+1$, is trivial. In fact, a real one-dimensional representation can only take the values $\pm 1$. A $1+1+1$ representation can then only be trivial or an overall sign change, thus providing no constraint on $m_E$, $m_\nu$. A less trivial example is $(A,0,0)$ (charged leptons) and $(a,b,b)$ (neutrinos). The charged lepton mass pattern requires $\rep_l$ to contain a one dimensional irrep corresponding to the non-vanishing mass and a possibly reducible two-dimensional representation corresponding to the two vanishing masses. The neutrino mass pattern requires a one dimensional real irrep, ``1'', together with either a two dimensional real irrep, ``2'', or the sum of a one dimensional complex representation and its conjugated, ``$\mathbf{1+\overline{1}}$''. We therefore have either $\rep_l \sim 1 + 2$ or $\rep_l \sim 1 + \mathbf{1+\overline{1}}$. In the first case, the irrep ``1'' must correspond to the non-zero charged lepton mass (the tau mass) and ``2'' must correspond to the two zero charged lepton masses (electron and muon masses). The representation $\rep_{e^c}$ must then be in the form $1+r$, where $r$ is a possibly reducible representation not equivalent to the irrep ``2''. In the second case, the irrep in $\rep_l$ corresponding to the tau mass can either be the real one or one of the complex ones ($\mathbf{1}$, without loss of generality). The forms of $\rep_{e^c}$ shown in table~\ref{tab:irreppattern1} follows. As a final example, consider the case in which the three neutrino masses are degenerate. The only possibility is that $\rep_l$ be a three dimensional real irrep. However, if that was the case, the three charged lepton masses would be forced to be degenerate, which is not a viable mass pattern (unless the masses are all vanishing, a case considered in section~\ref{sec:breaking}). There are therefore no possible groups and representations realising such a case in the symmetric limit. All the other cases in tables~\ref{tab:irreppattern1}, \ref{tab:irreppattern2} can be analysed in similar ways. 

It is now evident that the results in tables~\ref{tab:irreppattern1}, \ref{tab:irreppattern2} depend on the flavour group $G$ and on its representations $\rep_l$, $\rep_{e^c}$ on the leptons only through the structure of the decomposition of $\rep_l$, $\rep_{e^c}$ into irreducible components, and more precisely only on i) the dimensions of the irreps (the number denoting them in the table), ii) the possible equivalence or conjugation of the different components (conjugation is denoted by a bar over the representation, inequivalent irreps are distinguished by primes), and iii) whether the representation is complex/pseudoreal (boldface) or real (plain). The results show in particular that ($m^{(0)}_E \neq 0$, $m^{(0)}_\nu \neq 0$), i) the patterns with three degenerate non-vanishing neutrinos in the symmetric limit cannot be forced by any flavour group; ii) dimension 3 irreps are not involved in forcing any of the mass patterns we considered; iii) dimension 2 irreps can be contained in $\rep_{e^c}$ if, in the symmetric limit, $m_e = m_\mu = 0$; in $\rep_l$ if, in addition to that, $m_{\nu_1} = m_{\nu_2}$; iv) pseudoreal irreps can only play a role in $\rep_{e^c}$ if, in the symmetric limit, $m_e = m_\mu = 0$; in $\rep_l$ if, in addition to that, $m_{\nu_1} = m_{\nu_2} = 0$. 

\begin{table}[h]
\[
\begin{array}{|cc|lllll|}
\multicolumn{2}{c}{\text{lepton masses}} & \multicolumn{5}{c}{\text{decompositions of $\rep_l$ and $\rep_{e^c}$}}  \\[1mm]
\hline
(A00) & (aaa) &
\multicolumn{5}{c|}{\text{none}} \\
\hline
(A00) & (abb) &
\begin{array}{lll} \mathbf{1} & \overline{\mathbf{1}} & 1 \\ \overline{\mathbf{1}} & \multicolumn{2}{l}{r \nsupseteq 1,\mathbf{1}} \end{array} &
\begin{array}{lll} 1 & \mathbf{1} & \overline{\mathbf{1}} \\ 1 & \multicolumn{2}{l}{r \nsupseteq \overline{\mathbf{1}}, \mathbf{1}} \end{array} &
\begin{array}{lll} 1 & 2 &  \\ 1 &r \neq 2 \end{array} &
&
\\
\hline
(A00) & (0aa) &
\begin{array}{lll} \mathbf{1} & \mathbf{1}' & \overline{\mathbf{1}} \\ \overline{\mathbf{1}} & \multicolumn{2}{l}{r \nsupseteq \mathbf{1},\overline{\mathbf{1}'}} \end{array} &
\begin{array}{lll} \mathbf{1}' & \mathbf{1} & \overline{\mathbf{1}} \\ \overline{\mathbf{1}}' & \multicolumn{2}{l}{r \nsupseteq \mathbf{1},\overline{\mathbf{1}}} \end{array} &
\begin{array}{lll} \mathbf{1} & \mathbf{1} & \overline{\mathbf{1}} \\ \overline{\mathbf{1}} & \multicolumn{2}{l}{r \nsupseteq \mathbf{1},\overline{\mathbf{1}}} \end{array} &
\begin{array}{lll} \overline{\mathbf{1}} & \mathbf{1} & \mathbf{1} \\ \mathbf{1} & \multicolumn{2}{l}{r \nsupseteq \overline{\mathbf{1}}} \end{array} &
\begin{array}{lll} \mathbf{1} & 2 & \\ \overline{\mathbf{1}} & r \neq 2 \end{array} 
\\
\hline
(A00) & (a00) &
\begin{array}{lll} 1 & \mathbf{1} & \mathbf{1}' \\ 1 & \multicolumn{2}{l}{r \nsupseteq \overline{\mathbf{1}},\overline{\mathbf{1}'}} \end{array} &
\begin{array}{lll} \mathbf{1} & \mathbf{1}' & 1 \\ \overline{\mathbf{1}} & \multicolumn{2}{l}{r \nsupseteq 1,\overline{\mathbf{1}'}} \end{array} &
\begin{array}{lll} 1 & \mathbf{1} & \mathbf{1} \\ 1 & \multicolumn{2}{l}{r \nsupseteq \overline{\mathbf{1}}} \end{array} &
\begin{array}{lll} \mathbf{1} & \mathbf{1} & 1 \\ \overline{\mathbf{1}} & \multicolumn{2}{l}{r \nsupseteq 1, \overline{\mathbf{1}}} \end{array} &
\begin{array}{lll} 1 & \mathbf{2} & \\ 1 & r \nsupseteq \overline{\mathbf{2}} & \end{array} \\
\hline
(A00) & (abc) &
\begin{array}{lll} 1 & 1' & 1'' \\ 1 & \multicolumn{2}{l}{r \nsupseteq 1',1''} \end{array} &
\begin{array}{lll} 1 & 1 & 1' \\ 1 & \multicolumn{2}{l}{r \nsupseteq 1,1'} \end{array} &
\begin{array}{lll} 1' & 1 & 1 \\ 1' & \multicolumn{2}{l}{r \nsupseteq 1} \end{array} &
\begin{array}{lll} 1 & 1 & 1 \\ 1 & \multicolumn{2}{l}{r \nsupseteq 1} \end{array} &
\\
\hline
(A00) & (ab0) &
\begin{array}{lll} 1 & 1' & \mathbf{1} \\ 1 & \multicolumn{2}{l}{r \nsupseteq 1',\overline{\mathbf{1}}} \end{array} &
\begin{array}{lll} \mathbf{1} & 1' & 1 \\ \overline{\mathbf{1}} & \multicolumn{2}{l}{r \nsupseteq 1,1'} \end{array} &
\begin{array}{lll} \mathbf{1} & 1 & 1 \\ \overline{\mathbf{1}} & \multicolumn{2}{l}{r \nsupseteq 1} \end{array} &
\begin{array}{lll} 1 & 1 & \mathbf{1} \\ 1 & \multicolumn{2}{l}{r \nsupseteq 1,\overline{\mathbf{1}}} \end{array} &
\\
\hline
\end{array}
\]
\caption{Possible decompositions of $\rep_l$ (above) and $\rep_{e^c}$ (below) into irreducible components (part I). Each line corresponds to a combination of the charged lepton and neutrino mass patterns in the first two columns of \tab{masses}. Only the charged lepton pattern $(A00)$, which does not require hierarchies among non-zero entries, is considered here. Irreps are denoted by their dimensions. Boldface fonts denote complex or pseudoreal (if 2-dimensional) representations, regular fonts denote real representations. Primes are used to distinguish inequivalent representations, and in the case of complex representations $\mathbf{1}'$ is supposed to be different from both $\mathbf{1}$ and $\overline{\mathbf{1}}$.  ``$r$'' denotes a generic, possibly reducible representation, different from or not including the specified irreps, as indicated.}
\label{tab:irreppattern1}
\end{table}

\begin{table}[h]
\[
\begin{array}{|cc|lllll|}
\multicolumn{2}{c}{\text{lepton masses}} & \multicolumn{5}{c}{\text{decompositions of $\rep_l$ and $\rep_{e^c}$}}  \\[1mm]
\hline
(ABC) & (aaa) &
\multicolumn{5}{c|}{\text{none}} \\
\hline
(ABC) & (abb) &
\begin{array}{lll} 1 & \mathbf{1} & \overline{\mathbf{1}} \\ 1 & \overline{\mathbf{1}} & \mathbf{1} \end{array} &
&
&
&
\\
\hline
(ABC) & (0aa) &
\begin{array}{lll} \mathbf{1} & \mathbf{1}' & \overline{\mathbf{1}} \\ \overline{\mathbf{1}} & \overline{\mathbf{1}'} & \mathbf{1} \end{array} &
\begin{array}{lll} \mathbf{1} & \mathbf{1} & \overline{\mathbf{1}} \\ \overline{\mathbf{1}} & \overline{\mathbf{1}} & \mathbf{1} \end{array} &
&
&
\\
\hline
(ABC) & (a00) &
\begin{array}{lll} 1 & \mathbf{1} & \mathbf{1}' \\ 1 & \overline{\mathbf{1}} & \overline{\mathbf{1}'} \end{array} &
\begin{array}{lll} 1 & \mathbf{1} & \mathbf{1} \\ 1 & \overline{\mathbf{1}} & \overline{\mathbf{1}} \end{array} &
&
&
\\
\hline
(ABC) & (abc) &
\begin{array}{lll} 1 & 1' & 1'' \\ 1 & 1' & 1'' \end{array} &
\begin{array}{lll} 1 & 1 & 1' \\ 1 & 1 & 1' \end{array} &
\begin{array}{lll} 1 & 1 & 1 \\ 1 & 1 & 1 \end{array} &
&
\\
\hline
(ABC) & (ab0) &
\begin{array}{lll} 1 & 1' & \mathbf{1} \\ 1 & 1' & \overline{\mathbf{1}} \end{array} &
\begin{array}{lll} 1 & 1 & \mathbf{1} \\ 1 & 1 & \overline{\mathbf{1}} \end{array} &
&
&
\\
\hline
\hline
(AB0) & (aaa) &
\multicolumn{5}{c|}{\text{none}} \\
\hline
(AB0) & (abb) &
\begin{array}{lll} \mathbf{1} & \overline{\mathbf{1}} & 1 \\ \overline{\mathbf{1}} & \mathbf{1} & r\neq 1 \end{array} &
\begin{array}{lll} 1 & \mathbf{1} & \overline{\mathbf{1}} \\ 1 & \overline{\mathbf{1}} & r\neq \mathbf{1} \end{array} &
&
&
\\
\hline
(AB0) & (0aa) &
\begin{array}{lll} \overline{\mathbf{1}} & \mathbf{1} & \mathbf{1}'  \\ \mathbf{1} & \overline{\mathbf{1}} & r \neq \overline{\mathbf{1}'} \end{array} &
\begin{array}{lll} \mathbf{1} & \mathbf{1}' & \overline{\mathbf{1}} \\ \overline{\mathbf{1}} & \overline{\mathbf{1}'} & r\neq \mathbf{1} \end{array} &
\begin{array}{lll} \mathbf{1} & \mathbf{1} & \overline{\mathbf{1}} \\ \overline{\mathbf{1}} & \overline{\mathbf{1}} & r \neq \mathbf{1} \end{array} &
\begin{array}{lll} \overline{\mathbf{1}} & \mathbf{1} & \mathbf{1} \\ \mathbf{1} & \overline{\mathbf{1}} & r \neq \overline{\mathbf{1}} \end{array} &
\\
\hline
(AB0) & (a00) &
\begin{array}{lll} 1 & \mathbf{1} & \mathbf{1}' \\ 1 & \overline{\mathbf{1}} & r\neq \overline{\mathbf{1}'} \end{array} &
\begin{array}{lll} \mathbf{1} & \mathbf{1}' & 1 \\ \overline{\mathbf{1}} & \overline{\mathbf{1}'} & r \neq 1 \end{array} &
\begin{array}{lll} 1 & \mathbf{1} & \mathbf{1} \\ 1 & \overline{\mathbf{1}} & r \neq \overline{\mathbf{1}} \end{array} &
\begin{array}{lll} \mathbf{1} & \mathbf{1} & 1 \\ \overline{\mathbf{1}} & \overline{\mathbf{1}} & r \neq 1\end{array} &
\\
\hline
(AB0) & (abc) &
\begin{array}{lll} 1 & 1' & 1'' \\ 1 & 1' & r \neq 1'' \end{array} &
\begin{array}{lll} 1 & 1 & 1' \\ 1 & 1 & r \neq 1' \end{array} &
\begin{array}{lll} 1' & 1 & 1 \\ 1' & 1 & r \neq 1 \end{array} &
\begin{array}{lll} 1 & 1 & 1 \\ 1 & 1 & r \neq 1 \end{array} &
\\
\hline
(AB0) & (ab0) &
\begin{array}{lll} 1 & 1' & \mathbf{1} \\ 1 & 1' & r\neq \overline{\mathbf{1}} \end{array} &
\begin{array}{lll} 1 & \mathbf{1} & 1' \\ 1 & \overline{\mathbf{1}} & r\neq 1' \end{array} &
\begin{array}{lll} 1 & 1 & \mathbf{1} \\ 1 & 1 & r\neq \overline{\mathbf{1}} \end{array} &
\begin{array}{lll} \mathbf{1} & 1 & 1 \\ \overline{\mathbf{1}} & 1 & r\neq 1 \end{array} &
\\
\hline
\end{array}
\]
\caption{Possible decompositions of $\rep_l$ (above) and $\rep_{e^c}$ (below) into irreducible components (part II). Each line corresponds to a combination of the charged lepton and neutrino mass patterns in the first two lines of \tab{masses}. The charged lepton patterns $(ABC)$ and $(AB0)$ are considered here, which require hierarchies among the non-zero entries. Irreps are denoted by their dimensions. Boldface fonts denote complex representations, regular fonts denote real representations. Primes are used to distinguish inequivalent representations, and in the case of  complex representations $\mathbf{1}'$ is supposed to be different from both $\mathbf{1}$ and $\overline{\mathbf{1}}$.  ``$r$'' denotes a generic, possibly reducible representation, different from or not including the specified irreps, as indicated.}
\label{tab:irreppattern2}
\end{table}

\subsection{Accounting for lepton mixings}
\label{sec:mixings}

We have found so far the possible irrep decompositions leading, in the symmetric limit, to a reasonable approximation for the lepton masses. We now want to select those among them that also lead to a reasonable approximation for the PMNS matrix. As we will see, the form of the PMNS matrix only depends on the structure of the irrep decompositions, and can be determined in terms of the latter with simple rules that do not require the explicit construction of the mass matrices nor their diagonalisation. We will present in this section the results and leave the proofs to the appendix~\ref{sec:proofs}. 

The form of the PMNS matrix $\U$ associated to a given irrep decompositions of $\rep_l$ and $\rep_{e^c}$ in the symmetric limit, is 
\begin{equation}
\U = H^{\phantom{\dagger}}_E P^{\phantom{\dagger}}_E V D^{-1} P_\nu^{-1} H_\nu^{-1} \, .
\label{eq:mostgeneral}
\end{equation}
The contributions to $\U$ on the right hand side have different origins and different physical meanings. Each of them can be obtained without the need of writing explicitly nor diagonalising the lepton mass matrices, with the following rules. 
\begin{itemize}
\item
First, it is useful to order the irreps in such a way that those in $\rep_l$, $\rep_{e^c}$ that are conjugated to each other appear last and in the same position in the list. This way the vanishing charged lepton masses will appear first in the list of eigenvalues. For example, in one of the cases in table~\ref{tab:irreppattern2}, we could have $\rep_l =  \mathbf{1} + \mathbf{1} + \overline{\mathbf{1}}$, $\rep_{e^c} = (r \neq \overline{\mathbf{1}}) + \overline{\mathbf{1}} + \mathbf{1}$. Correspondingly, we write a list of generic charged lepton eigenvalues with the non-vanishing eigenvalues corresponding to the conjugated representations. In the example above, the list would be $(0,B,A)$.\footnote{Note that in the tables, for convenience, the three families appear in inverse order: (3,2,1).}
\item
$V$ is a generic unitary transformation commuting with $\rep_l$, with $\ord{1}$ entries. Its origin is associated to the presence of equivalent copies of the same irrep type in the decomposition of $\rep_l$. If all the irrep components are inequivalent, $V$ is trivial. For example, if $\rep_l  = \mathbf{1} + \mathbf{1} + \overline{\mathbf{1}}$, $V$ is a $2 \times 2$ unitary transformation in the 12 block. 
\item
$D$ is associated to the possible presence of a Dirac sub-structure in the neutrino mass matrix, and it originates from the presence of complex conjugated irreps within the decomposition of $\rep_l$. In the three neutrino case, there are only two possibilities. Either $\rep_l$ does not contain pairs of complex conjugated irreps, in which case $D$ is trivial, $D_{ij} = \delta_{ij}$. Or there is one pair of one-dimensional complex conjugated representations, in the positions $i$ and $j$ in the list of irreps, in which case $D$ is a maximal $2\times 2$ rotation, 
\begin{equation}
D_2 = 
\frac{1}{\sqrt{2}}
\begin{pmatrix}
1 & 1 \\ - i & i 
\end{pmatrix} ,
\label{eq:D}
\end{equation}
embedded in the $ij$ block. The corresponding mass eigenvalues are degenerate (both positive due to the imaginary unit in $D_2$, contributing to the Majorana phases). Correspondingly, we write the list of neutrino eigenvalues as follows. If a pair of conjugated irreps is present in $\rep_l$ in the positions $i$ and $j$, we have two degenerate non-vanishing eigenvalues in the corresponding positions. We then have a non-vanishing eigenvalue in the position corresponding to each real representation. If the real irrep has dimension $d>1$, there will be $d$ degenerate eigenvalues. Finally, we have a vanishing eigenvalue corresponding to each unmatched complex representation. In the previous example, with $\rep_l  = \mathbf{1} + \mathbf{1} + \overline{\mathbf{1}}$, we can equivalently choose the two conjugated representations to be the ones in the positions $ij = 23$ or those in the positions $ij = 13$. Such a choice will determine the positions $i$ and $j$ of the corresponding two degenerate neutrino masses in the list of eigenvalues (before the reordering below). So if we choose $ij = 23$, we will have the $2\times 2$ block in \eq{D} embedded in the 23 block of the matrix $D$ and the list of neutrino eigenvalues will be in the form $(0aa)$. 
\item
The permutation matrices $P_E$ and $P_\nu$ are associated to the possible need of reordering the list of eigenvalues. Indeed, the list of eigenvalues obtained with the above rules is not necessarily in the standard ordering, required for a proper definition of the PMNS matrix. In the example we have considered, the list of charged lepton eigenvalues is $(0, B, A)$. The masses are in standard ordering if $B<A$. On the other hand, if $B>A$, the standard ordering is obtained by switching $A$ and $B$. Correspondingly, $P_E$ is either the identity or a permutation matrix switching $2\leftrightarrow 3$. As for the neutrinos, the list of eigenvalues is in the form $(0aa)$. The standard ordering requires the two degenerate eigenvalues to be in the first two positions. Therefore, $P_\nu$ is a permutation matrix moving the first entry in the third position. 
\item
Finally, the role of $H_e$, $H_\nu$ is to take into account possible ambiguities in the definition of the PMNS matrix in the symmetric limit. In the real world, all leptons  are non-degenerate and the PMNS matrix only has unphysical phase ambiguities, which do not need to be taken into account. When considering the symmetric limit, on the other hand, larger ambiguities can arise due  to degenerate, possibly vanishing, masses. In practice, $H_E$ is a generic unitary transformation mixing the massless charged leptons; and $H_\nu$ contains a generic unitary transformation mixing the massless neutrinos and a generic orthogonal transformation mixing degenerate massive neutrinos (it turns out, however, that the latter can be ignored if the degeneracy is due to a Dirac structure, in which case it can be reabsorbed into a phase redefinition of $V$). As discussed in the Appendix, the $H_e$, $H_\nu$ contributions to the PMNS matrix have a different physical nature than the previous ones. They are unphysical, and undetermined, in the symmetric limit. However, they become physical (up to diagonal phases) after symmetry breaking effects split the degenerate mass eigenstates. Depending on the specific form of the symmetry breaking effects, $H_e$ and $H_\nu$ can end up being be large, small, or zero (unlike the previous contributions, which are determined by the non-zero entries and are large in the absence of accidental correlations~\cite{Domcke:2016mzc}). 
\end{itemize}

With the above rules, we can determine the form of the PMNS matrix associated to each irrep pattern in tables~\ref{tab:irreppattern1}, \ref{tab:irreppattern2} and select the cases leading to a PMNS matrix with no zeros or a zero in the 13 position. The results are illustrated in table~\ref{tab:PMNS}. 

\begin{table}[h]
\[
\begin{array}{|l|c|c|cccccc|c|c|}
\hline
\text{irreps} & \text{masses} & \text{$\nu$ hierarchy} & H_E & P_E & V & D & P_\nu & H_\nu & \U  & \text{zeros}  \\[1mm]
\hline
\begin{array}{lll} 1 & 1 & 1 \\ 1 & \multicolumn{2}{l}{r \nsupseteq 1} \end{array} &
\begin{array}{c} (A00) \\ (abc) \end{array} &
\text{NH or IH} &
 &
 &
V &
 &
 &
 &
V &
\text{none}
 \\
\hline
\begin{array}{lll} \mathbf{1} & \mathbf{1} & \overline{\mathbf{1}} \\ \overline{\mathbf{1}} & \multicolumn{2}{l}{r \nsupseteq \mathbf{1},\overline{\mathbf{1}}} \end{array} &
\begin{array}{c} (A00) \\ (0aa) \end{array} &
\text{IH} &
H^E_{12} &
 &
V_{23} &
D_{12} &
 &
 &
H^E_{12} V_{23} D^{-1}_{12} &
\text{none} \, (\mathbf{13}) 
 \\
\hline
\hline
\begin{array}{lll} 1 & 1 & 1 \\ 1 & 1 & r \neq 1 \end{array} &
\begin{array}{c} (AB0) \\ (abc) \end{array} &
\text{NH or IH} &
&
 &
V &
 &
 &
 &
V &
\text{none} 
 \\
\hline
\begin{array}{lll} \mathbf{1} & \mathbf{1} & \overline{\mathbf{1}} \\ \overline{\mathbf{1}} & \overline{\mathbf{1}} & r \neq \mathbf{1} \end{array} &
\begin{array}{c} (AB0) \\ (0aa) \end{array} &
\text{IH} &
 &
 &
V_{23} &
D_{12} &
 &
 &
V_{23} D^{-1}_{12} &
\mathbf{13} 
 \\
\hline
\begin{array}{lll} 1 & 1 & 1 \\ 1 & 1 & 1 \end{array} &
\begin{array}{c} (ABC) \\ (abc) \end{array} &
\text{NH or IH} &
 &
 &
V &
 &
 &
 &
V &
\text{none}
\\
\hline
\begin{array}{lll} \mathbf{1} & \mathbf{1} & \overline{\mathbf{1}} \\
\overline{\mathbf{1}} & \overline{\mathbf{1}} & \mathbf{1} \end{array} &
\begin{array}{c} (ABC) \\ (0aa) \end{array} &
\text{IH} &
 &
P_E &
V_{23} &
D_{12} &
 &
 &
P_E V_{23} D^{-1}_{12} &
\mathbf{13}, 23, 33 
 \\
\hline
\end{array}
\]
\caption{Irrep decompositions giving rise to a PMNS matrix with no zeros or a single zero possibly in the 13 entry. The first column shows the decomposition of $\rep_l$ and $\rep_{e^c}$, one above the other. Only real and complex irreps appear. The second column shows the corresponding pattern of charged lepton and neutrino masses in the symmetric limit, one above the other, and the third the neutrino hierarchy type, normal (NH) or inverted (IH). The individual contributions to the PMNS matrix are then shown. A matrix with no further specification is generic (e.g.\ $P$ denotes a generic permutation, $V$ a generic unitary matrix). $D_{ij}$ denotes a $\pi/4$ rotation in the generic form in \eq{D} acting in the sector $ij$. If no information on a certain factor is given, that factor is irrelevant (for example because diagonal or because it can be reabsorbed in another factor). The presence and position of a zero in the PMNS matrix in the symmetric limit is specified in the last column.}
\label{tab:PMNS}
\end{table}

As shown, there is a limited number of cases leading, in the symmetric limit, to lepton observables close to what observed. Each case corresponds to a certain decomposition of the flavour representations in terms of real and complex, equivalent and inequivalent representations of given dimension. Each  pattern may correspond to different flavour groups and representations, provided that the decomposition of the representation on the leptons follows that pattern. The allowed patterns contain one-dimensional irreps only. Pseudoreal representations do not play a role. 

Three out of the six cases in the table are partially trivial. Those are the cases in which $\rep_l \sim 1 + 1 + 1$, for which the representation on the lepton doublets is either the  identity representation or an overall sign change. In such a case, the neutrino mass matrix is not constrained at all, and the neutrino masses and PMNS matrix are expected to be completely generic. In particular, the relative smallness of $|(\U)_{13}|$ is accidental. We are in the presence of ``anarchical'' neutrinos~\cite{Hall:1999sn,Haba:2000be}. The only constraints that can be obtained are on the charged lepton masses, through the interplay of the trivial $\rep_l$ with a non-trivial $\rep_{e^c}$. 

The other three cases provide non-trivial constraints on neutrino masses and mixings. An important result is that they all correspond to inverted neutrino hierarchy, and specifically to two degenerate and one vanishing neutrino mass in the symmetric limit. Therefore, if the present hint favouring a normal hierarchy were confirmed, we would conclude, within our assumptions, that either the flavour model is not predictive at all in the neutrino sector, or the symmetric limit does not provide an approximate description of lepton masses and mixings. In the latter case, we might have to resort to a caveat in our assumptions (see conclusions) or to the case where all charged lepton or all neutrino masses vanish in the symmetric limit (last column of table~\ref{tab:masses}), and symmetry breaking effects are crucial to understand even the basic features of lepton mixing. 

Table~\ref{tab:PMNS} is divided in two parts. In the first part, the hierarchy of the charged lepton masses is naturally accommodated by the vanishing of the two lighter masses in the symmetric limit, in agreement with the principle that hierarchies should be explained by the flavour model. In the second part, hierarchies not accounted for by the flavour theory have to be invoked among the non-zero entries in order to account for the structure of charged lepton masses. The second case in the first part of the table is special, as the size of the 13 element of the PMNS matrix is determined by the rotation $H^E_{12}$, which is not physical in the symmetric limit, and will be fixed by the symmetry breaking effects generating the muon mass. Depending on the structure of those effects, the size of $(\U)_{13}$ can end up being large, small, or zero. Finally, note that since the parameters entering all the mixing matrices in \tab{PMNS} except $D$ are generic, a specific value of a mixing angle can be obtained only when the matrix $D$ is involved. As the table shows, $D$ can only play a role in the 12 mixing, in agreement with earlier specific results~\cite{Altarelli:2005yp}. 

In the next subsection, we shortly illustrate a few examples of specific flavour groups and representations corresponding to the patterns in table~\ref{tab:PMNS}.

\subsection{Examples}
\label{sec:groups}

The results above have been obtained without the need to specify the form of the lepton mass matrices, as they directly followed from the structure of the irrep decompositions. Moreover, there was no need to specify a flavour group or its representation on leptons, as the results hold for any group, of any type, as long as the decompositions of its representations have the structure shown in the tables. In the following, for completeness and as proofs of existence, we will provide examples, in some cases well known, of explicit realisations of the three cases in \tab{PMNS} leading to a PMNS matrix with a (possible) zero in the 13 position in the symmetric limit. All of them require a continuous or discrete symmetry group $G$ with a complex one-dimensional representation $\mathbf{1}$, and a representation on the leptons doublets decomposing as $\rep_l = \mathbf{1} + \mathbf{1} + \overline{\mathbf{1}}$. 

\subsubsection*{$\rep_l = \mathbf{1} + \mathbf{1} + \overline{\mathbf{1}}$, $\rep_{e^c} = \overline{\mathbf{1}} + (r \nsupseteq \mathbf{1},\overline{\mathbf{1}})$}

In this case, corresponding to the second row in \tab{PMNS}, the representation on the lepton singlets decomposes into a copy of $\overline{\mathbf{1}}$ and a (possibly reducible) two dimensional representation $r$ whose only requirement is not to contain either $\mathbf{1}$ or $\overline{\mathbf{1}}$ ($r$ could be for example the trivial representation). In the symmetric limit, two charged leptons are forced to be massless, which explains the suppression of the electron and muon mass compared to the tau mass, and the neutrino spectrum turns out to be inverted hierarchical, with $m_3 = 0$ and $m_1=m_2$. With the notations used in \tab{PMNS}, we thus have
\begin{equation}
(m_\tau,m_\mu,m_e) = (A,0,0),
\qquad
(m_{\nu_3},m_{\nu_2},m_{\nu_1}) = (0,a, a).
\label{eq:m1}
\end{equation}
A non-vanishing value of $m_e, m_\mu$ must then be generated by the symmetry breaking effects, which will also give $m_{3} \ll m_1 \approx m_2$. 

The PMNS matrix does not necessarily have a zero, as it is obtained from the combination of 3 rotations: $V_{23}$, the $\ord{1}$ rotation in the 23 sector commuting with $\rep_l$; a maximal 12 rotation $D_{12}$ associated with the Dirac substructure in $m_\nu$ forced by $\rep_l$; and a rotation $H^E_{12}$ in the 12 sector, associated to the degeneracy of the first two charged leptons and not determined in the symmetric limit. The latter is  fixed by the symmetry breaking effects generating the muon and electron masses. If the $H^E_{12}$ is large,  the PMNS matrix is expected not to have any small entry. On the other hand, in the light of the hierarchy $m_e \ll m_\mu$, one can expect $H^E_{12}$, and consequently $(\U)_{13}$, to be relatively small~\cite{Frampton:2004ud,Romanino:2004ww,King:2005bj,Antusch:2005kw,Hochmuth:2007wq,BhupalDev:2011gi,Dev:2011bd,Marzocca:2011dh,Altarelli:2012ss,Marzocca:2013cr,Gollu:2013yla,Marzocca:2014tga}. The PMNS matrix thus reads
\begin{equation}
\setlength\arraycolsep{5pt}
\U = H^E_{12} V_{23} D^{-1}_{12} = \begin{pmatrix}
X & X & ? \\
X & X & X \\
X & X & X
\end{pmatrix} ,
\label{eq:U1}
\end{equation}
where $X$ denotes a non-zero entry, not further constrained, and the size of the 13 entry depends on $H^E_{12}$, as discussed. The form of lepton mass matrices in the symmetric limit is
\begin{equation}
\setlength\arraycolsep{5pt}
m_E = \begin{pmatrix}
 & & \\
& & \\
\phantom{X} & X & X
\end{pmatrix},
\qquad
m_\nu = \begin{pmatrix}
& X & X \\
X & &  \\
X & &
\end{pmatrix}.
\label{eq:case3}
\end{equation}
It is easy to exhibit an example of a group $G$ and representations $\rep_l$, $\rep_{e^c}$ with a decomposition in irreps as above. An easy choice is $G = U(1)$, with $\omega\in U(1)$ represented by 
\begin{equation}
\label{eq:rep1}
\setlength\arraycolsep{5pt}
\rep_l(\omega) = 
\begin{pmatrix}
\omega^* & & \\
& \omega & \\
& & \omega
\end{pmatrix} ,
\quad
\rep_{e^c}(\omega) = 
\begin{pmatrix}
\omega^q & & \\
& \omega^p & \\
& & \omega^*
\end{pmatrix} ,
\end{equation}
where $p,q\neq \pm1$. For example, one can choose $p = q = 0$ (trivial representation). A minimal possibility involving a discrete group is $G = \mathbf{Z}_3$, with the same representation of $\omega \in \mathbf{Z}_3$ and $p = q = 0$ as the only possible choice. Any other discrete subgroup of $U(1)$, different from $\mathbf{Z}_2$ would of course also work. It is also possible to realise this case by using the one dimensional representations of non-abelian discrete groups, such as $A_4$ for example. 

\subsubsection*{$\rep_l = \mathbf{1} + \mathbf{1} + \overline{\mathbf{1}}$, $\rep_{e^c} = \overline{\mathbf{1}} + \overline{\mathbf{1}} + (r \neq \mathbf{1})$}

In this case, corresponding to the fourth row in \tab{PMNS}, the representation on the lepton singlets decomposes into two copies of $\overline{\mathbf{1}}$ and a one dimensional representation $r$ inequivalent to $\mathbf{1}$. In the symmetric limit, one charged lepton is forced to be massless, which explains the suppression of the electron mass compared to the muon and tau mass, but not the hierarchy $m_\mu \ll m_\tau$, and the neutrino spectrum turns out to be inverted hierarchical as before,
\begin{equation}
(m_\tau,m_\mu,m_e) = (A,B,0),
\qquad
(m_{\nu_3},m_{\nu_2},m_{\nu_1}) = (0,a, a).
\label{eq:m2}
\end{equation}

The PMNS matrix contains a zero, unambiguously positioned in the 13 entry. It is obtained from the combination of 2 rotations: $V_{23}$, the $\ord{1}$ rotation in the 23 sector commuting with $\rep_l$, and a maximal 12 rotation $D_{12}$. Unlike the previous case, the form of the PMNS matrix is determined in the symmetric limit up to phase ambiguities only. The forms of the PMNS matrix and the of the lepton mass matrices in the symmetric limit are
\begin{equation}
\setlength\arraycolsep{5pt}
\U = \begin{pmatrix}
X & X & 0 \\
X & X & X \\
X & X & X
\end{pmatrix} , \quad
\setlength\arraycolsep{5pt}
m_E = \begin{pmatrix}
 & & \\
& X & X \\
\phantom{X} & X & X
\end{pmatrix},
\quad
m_\nu = \begin{pmatrix}
& X & X \\
X & &  \\
X & &
\end{pmatrix} .
\label{eq:U2}
\end{equation}
A simple implementation of this case can be obtained from the previous one by modifying the way the group acts on $\mu^c$. For $G = U(1)$, we can in fact represent $\omega\in U(1)$ by 
\begin{equation}
\label{eq:rep2}
\setlength\arraycolsep{5pt}
\rep_l(\omega) = 
\begin{pmatrix}
\omega^* & & \\
& \omega & \\
& & \omega
\end{pmatrix} ,
\quad
\rep_{e^c}(\omega) = 
\begin{pmatrix}
\omega^q & & \\
& \omega^* & \\
& & \omega^*
\end{pmatrix} ,
\end{equation}
where $q\neq 1$, for example $q = 0$. As before, abelian or non-abelian discrete groups can also be used. 
 
\subsubsection*{$\rep_l = \mathbf{1} + \mathbf{1} + \overline{\mathbf{1}}$, $\rep_{e^c} = \overline{\mathbf{1}} + \overline{\mathbf{1}} + \mathbf{1}$}

This case, corresponding to the sixth row in \tab{PMNS}, has a particularly well known implementation: $G = U(1)$ acting on leptons according to their  $L_\tau + L_\mu - L_e$ charge~\cite{Petcov:1982ya,Barbieri:1998mq,Joshipura:1998kg,Mohapatra:1999zr,Petcov:2004rk,Altarelli:2005pj}. The disadvantage of this case is that, whatever is the implementation, none of the charged lepton hierarchies, $m_e \ll m_\mu \ll m_\tau$, is explained by the model. The neutrino spectrum is inverted hierarchical, as before, and with the notations used in \tab{PMNS} we have

\begin{equation}
(m_\tau,m_\mu,m_e) = (A,B,C),
\qquad
(m_{\nu_3},m_{\nu_2},m_{\nu_1}) = (0,a, a).
\label{eq:m3}
\end{equation}

Another disadvantage is that the PMNS matrix does contain a zero, but the model does not explain why it appears in the 13 entry, as in principle it could also appear in the 23 or 33 entry. This is because the permutation $P_E$ in \eq{mostgeneral}, sorting the charged leptons in the standard order, is generic in this case. In other words, the symmetry does force the eigenvalue positioned where the electron should be to be the lightest, and a viable symmetric limit for the PMNS matrix is obtained only in that case, i.e.\ when the smallest eigenvalue happens to correspond to the lepton transforming as $\overline{\mathbf{1}}$ under $\rep_l$. In such a case, the PMNS matrix and the lepton mass matrices in the symmetric limit are in the form
\begin{equation}
\setlength\arraycolsep{5pt}
\U = \begin{pmatrix}
X & X & 0 \\
X & X & X \\
X & X & X
\end{pmatrix} , \quad
\setlength\arraycolsep{5pt}
m_E = \begin{pmatrix}
X & & \\
& X & X \\
\phantom{X} & X & X
\end{pmatrix},
\quad
m_\nu = \begin{pmatrix}
& X & X \\
X & &  \\
X & &
\end{pmatrix} .
\label{eq:U3}
\end{equation}

\section{Lepton mixing from symmetry breaking effects} 
\label{sec:breaking}

We will now consider the cases in which all neutrinos or all charged lepton masses vanish in the symmetric limit ($m^{(0)}_E = 0$ or $m^{(0)}_\nu = 0$ in \eq{main}), i.e.\ the cases associated to the last column in \tab{masses}. In such cases, the sole knowledge of the flavour group and its representation is not sufficient to account for any of the features of lepton mixing, as the PMNS matrix is completely undetermined (unphysical) in the symmetric limit, with its final form fully depending on the symmetry breaking effects. 

As symmetry breaking effects are now central, let us consider not only the flavour group $G$ and its representations on the leptons, here denoted by $\rep^G_l$ and $\rep^G_{e^c}$, but also the residual group $H$ to which $G$ is spontaneously broken, and its representations on leptons $\rep^H_l$ and $\rep^H_{e^c}$, which are simply the restriction to $H$ of $\rep^G_l$ and $\rep^G_{e^c}$. If $G$ is fully broken, the residual group $H$ only contains the identity, and its representations are trivial. Symmetry breaking can take place in more than one step, $G \to H_1 \to \ldots \to H_n$, associated to different scales. In such a case, our results will correspond to the first step of the breaking chain, $H = H_1$, and the corresponding breaking effects will only provide a leading order prediction for the lepton observables, as the contribution of the subsequent steps may be needed to precisely fit them. 

We want to characterise the forms of $\rep^G_l$ and $\rep^G_{e^c}$ and $\rep^H_l$ and $\rep^H_{e^c}$ leading, once $G$ is broken (but $H$ is not), to a pattern of lepton masses and mixing not far from what observed. 

Such a problem does not admit a general answer as simple as the one obtained in the previous section. The reason is that the final pattern of lepton observables does not only depend on $G$, $H$, $\rep^G$, $\rep^H$, but it also depends on the specific spectrum of flavons and their vevs (and the scalar potentials determining the vevs). On the other hand, it turns out that a simple answer can be obtained if the following (non-trivial) hypothesis holds: the symmetry breaking corrections, $m^{(1)}_E$, $m^{(1)}_\nu$ in \eq{main}, have the most general form allowed by the residual symmetry $H$, with all non-vanishing entries of the same order. Needless to say, neither neutrino nor charged lepton masses should identically vanish after symmetry breaking. In such a case, it turns out that the formalism developed and the results obtained in the previous sections on the possible structures of $\rep^G$ can be simply reinterpreted in terms of the possible structures of $\rep^H$, as we will see below. 

The hypothesis we introduced is non-trivial. It amounts at assuming that the lepton observables only depend on the symmetry breaking pattern $G \to H$ and not on the specific breaking mechanism used. This is not the case in most models found in the literature, in which the flavour structure is rather associated to the specific choice of the flavon spectrum, to their coupling to the leptons, and to the form of their vevs. This is the case for example in models where the residual symmetry $H$ is different in the neutrino and charged lepton sectors; and even in the case of U(1) models, in which $H = \{1\}$, all entries are allowed by $H$, but they typically turn out to be of different sizes, depending on how many powers of the flavons are needed to generated them. Still, the results we will get under the above hypothesis are useful for a complete assessment of the importance of a detailed knowledge of the symmetry breaking mechanism.

Let us motivate the result mentioned above. Suppose, as we do, that $G$ is spontaneously broken to $H$, that either the charged lepton or the neutrino masses (not both) vanish in the $G$-symmetric limit, and that, after spontaneous breaking, we obtain a mass pattern close to what observed, i.e.\ in one of the forms listed in the first two columns of \tab{masses}. The knowledge of the mass pattern after symmetry breaking allows us to constrain $\rep^H$. The possible structures of the irrep decomposition of the representation $\rep^H$ are in fact listed, for each mass pattern, in tables~\ref{tab:irreppattern1}, \ref{tab:irreppattern2}, where $\rep_l$ and $\rep_{e^c}$ should now be interpreted as $\rep^H_l$ and $\rep^H_{e^c}$. The group $G$ plays no role at this point. A further constraint comes from the requirement that the PMNS be also close to what observed after symmetry breaking. In order to find the form of the PMNS matrix associated to a given breaking pattern, we can proceed as in the appendix. We then find that the form of the PMNS matrix again depends on $\rep^H$ only, and its structure is still given by \eq{mostgeneral}, with the form of each factor dictated by the same rules given in that section, where $\rep_l$ and $\rep_{e^c}$ should now be interpreted as $\rep^H_l$ and $\rep^H_{e^c}$. The group $G$ again plays no role.\footnote{The only possible role of $G$ is in the determination of $V_e$, $V_\nu$ in \eq{V}, obtained by the diagonalisation of $m_{E,r}$, $m_{\nu,r}$ in eqs.~(\ref{eq:blockE},\ref{eq:blocknu},\ref{eq:blockps}), which now include symmetry breaking effects. In one of the two matrices, say $m_{E,r}$ for definiteness, two scales now enter, the scale of $m^{(0)}_E$ and the scale of $m^{(1)}_\nu$ (while in the neutrino sector $m^{(0)}_\nu = 0$ and only one scale appears). In such a case $V_{e,r}$ may not be a generic matrix with $\ord{1}$ entries, it could for example contain small mixing angles. On the other hand, only one scale, that of $m^{(1)}_\nu$, enters $m_{\nu,r}$, so that $V_{\nu,r}$ is still a generic matrix with $\ord{1}$ entries. As $V$ is the combination of $V_e$, and $V_\nu$, $V$ will be also a generic matrix with $\ord{1}$ entries, whatever is the form of $V_e$.}  We conclude that the structure of the irrep decomposition of $\rep^H$ must be one of those in \tab{PMNS}, where once again $\rep_l$ and $\rep_{e^c}$ should be interpreted as $\rep^H_l$ and $\rep^H_{e^c}$, and the mass pattern and PMNS matrix after symmetry breaking can only be in the forms shown in that table. 

The presence of an unbroken, $G$-symmetric phase played no role in constraining the form of $\rep^H$, nor in determining the form of the PMNS matrix. On the other hand, it can play a useful role in providing hierarchies among lepton masses, in particular within the more hierarchical charged lepton masses. We have in fact now two scales available in the sector, let us say the charged lepton one for definiteness, where the masses do not vanish in the symmetric limit: the scale of the non-vanishing entries in $m^{(0)}_E$, allowed by $G$; and the lower scale of the non-vanishing entries in $m^{(1)}_E$, allowed by $H$ but not by $G$. We can then use the ratio between those two scales to account for the hierarchy between the tau and muon masses. Therefore, while in section \ref{sec:mixings} we focussed only on the first two lines in \tab{PMNS}, as in the other part of the table the needed hierarchies were not accounted for, now all the first four lines are on the same footing. The hierarchy needed between $A$ and $B$ in the cases in which the charged lepton masses are in the form $(A,B,0)$ can in fact be provided by the two scales above. On the other hand, the last two lines, corresponding to the $(A,B,C)$ pattern, are still not on the same footing, as they require two hierarchies to be explained. 

Let us discuss in greater detail how the available hierarchy can enter the results in \tab{PMNS}. Let us first explicitly list the possible mass patterns in the $G$-symmetric limit. There are two cases. Either the neutrino masses all vanish, in which case the charged lepton masses are in the form $(A,0,0)$ (we discard $(A,B,0)$ and $(A,B,C)$ at this level as in the symmetric limit there is only one scale); or the charged lepton masses all vanish, in which cases neutrino masses are in one of the forms $(a,a,a)$, $(a,b,b)$, $(a,b,c)$, $(0,a,a)$, $(a,b,0)$, $(a,0,0)$. Let us now switch on the symmetry breaking effects. The charged and neutral lepton masses will then get additional contributions from $m^{(1)}_E$, $m^{(1)}_\nu$, which we can denote as proportional to a parameter $\epsilon$. In the sector in which $m^{(0)} \neq 0$, the $\epsilon$ parameter represents the ratio of the two scales, $m^{(1)}$ and $m^{(0)}$.\footnote{In principle the correction to the masses could be proportional to higher powers of $\epsilon$, but it turns out that this it not the case, under our hypotheses.} In \tab{GtoH} we show the lepton mass patterns that can be obtained, together with a viable PMNS matrix, taking into account the presence of the two scales. We discard the $(A,B,C)$ charged lepton pattern (last two lines in \tab{PMNS}) as it requires at least one unaccounted hierarchy. In \tab{GtoH}, the lepton mass pattern in the $G$-symmetric limit can be obtained by setting $\epsilon = 0$. The corresponding irrep decompositions of $\rep^G$ are shown, as well as the irrep decomposition of $\rep^H$ shaping the symmetry breaking corrections. We have checked that for each pair of irrep decompositions of $\rep^G$ and $\rep^H$ in the table corresponding to the same mass pattern there exists concrete examples of  the groups $G$ and $H$ and of the representations of $G$, $\rep^G_l$ and $\rep^G_{e^c}$, such that the decomposition of the latter under $H$ reproduces the chosen irrep decomposition of $\rep^H$. 

\begin{table}[h]
\[
\begin{array}{|c|c|l|l|c|c|}
\hline
\text{masses} & \text{hierarchy} &  \multicolumn{1}{|c|}{ G \; \text{irreps} } &  H \; \text{irreps} & \U  & \text{zeros}  \\[1mm]
\hline
\begin{array}{c} (\,\makebox[\myem]{$\epsilon A$}\,\makebox[\myem]{$0$}\makebox[\myl]{$0$}) \\ (\makebox[\myl]{$a$}\,\makebox[\myem]{$b$}\makebox[\myl]{$c$}) \end{array} &
\text{NH or IH} &
\begin{array}{lll} 1 & 1 & 1 \\ \multicolumn{3}{l}{r \nsupseteq 1 }   \end{array} &
\begin{array}{lll} 1 & 1 & 1 \\ 1 & \multicolumn{2}{l}{r \nsupseteq 1} \end{array} &
V &
\text{none}
\\
\hline
\begin{array}{c} (\,\makebox[\myem]{$\epsilon A$}\,\makebox[\myem]{$0$}\makebox[\myem]{$0$}) \\ (\makebox[\myl]{$a$}\,\makebox[\myem]{$b$}\,\makebox[\myem]{$\epsilon c$}) \end{array} &
\text{NH or IH} &
\begin{array}{lll} 1 & 1 & \mathbf{1} \\ \multicolumn{3}{l}{r \nsupseteq 1, \overline{\mathbf{1}}}  \end{array} &
\begin{array}{lll} 1 & 1 & 1 \\ 1 & \multicolumn{2}{l}{r \nsupseteq 1} \end{array} &
V &
\text{none}
\\
\hline
\begin{array}{c} (\,\makebox[\myem]{$\epsilon A$}\,\makebox[\myem]{$0$}\makebox[\myem]{$0$}) \\ (\makebox[\myl]{$a$}\;\makebox[\myem]{$\epsilon b$}\;\makebox[\myem]{$\epsilon c$}) \end{array} &
\text{NH} &
\hspace{-1.5mm}\begin{array}{lll}
\begin{array}{lll} 1 & \mathbf{1} & \mathbf{1} \\  \multicolumn{3}{l}{r \nsupseteq 1,\overline{\mathbf{1}}} \end{array} &
\begin{array}{lll} 1 & \mathbf{1} & \mathbf{1}'\\  \multicolumn{3}{l}{r \nsupseteq 1,\overline{\mathbf{1}}, \overline{\mathbf{1'}}} \end{array} \vspace{1.2pt} &
\begin{array}{lll} 1 & \mathbf{2} &   \\ \multicolumn{3}{l}{r \nsupseteq 1,\overline{\mathbf{2}}} \end{array}
\end{array} &
\begin{array}{lll} 1 & 1 & 1 \\ 1 & \multicolumn{2}{l}{r \nsupseteq 1} \end{array} &
V &
\text{none}
\\
\hline
\begin{array}{c} (\makebox[\myl]{$A$}\,\makebox[\myem]{$0$}\makebox[\myl]{$0$}) \\ (\makebox[\myem]{$\epsilon a$}\;\makebox[\myem]{$\epsilon b$}\;\makebox[\myl]{$\epsilon c$}) \end{array} &
\text{NH or IH} &
\hspace{-1.5mm}
\begin{array}{lll}
\begin{array}{lll} \mathbf{1} & \mathbf{1} & \mathbf{1} \\ \overline{\mathbf{1}} & \multicolumn{2}{l}{r \nsupseteq \overline{\mathbf{1}}} \end{array} &
\begin{array}{lll} \mathbf{1} & \mathbf{1} & \mathbf{1'} \\ \overline{\mathbf{1}} & \multicolumn{2}{l}{r \nsupseteq \overline{\mathbf{1}}, \overline{\mathbf{1'}}} \end{array} \vspace{1.4pt}  &
\begin{array}{lll} \mathbf{1'} & \mathbf{1} & \mathbf{1} \\ \overline{\mathbf{1'}} & \multicolumn{2}{l}{r \nsupseteq \overline{\mathbf{1}}} \end{array} \vspace{1.2pt}  \\
\begin{array}{lll} \mathbf{1} & \mathbf{1'} & \mathbf{1''} \\ \overline{\mathbf{1}} & \multicolumn{2}{l}{r \nsupseteq \overline{\mathbf{1'}}, \overline{\mathbf{1''}}} \end{array} &
\begin{array}{lll} \mathbf{1} & \mathbf{2} &   \\ \overline{\mathbf{1}} & \multicolumn{2}{l}{r \neq \overline{\mathbf{2}}} \end{array} &
\end{array}&
\begin{array}{lll} 1 & 1 & 1 \\ 1 & \multicolumn{2}{l}{r \nsupseteq 1} \end{array} &
V &
\text{none}
\\
\hline
\begin{array}{c} (\makebox[\myl]{$A$}\,\makebox[\myem]{$0$}\makebox[\myl]{$0$}) \\ (0\;\makebox[\myem]{$\epsilon a$}\;\makebox[\myem]{$\epsilon a$}) \end{array} &
\text{IH} &
\hspace{-1.5mm}\begin{array}{lll}
\begin{array}{lll} \mathbf{1} & \mathbf{1} & \mathbf{1'} \\ \overline{\mathbf{1}} & \multicolumn{2}{l}{r \nsupseteq \overline{\mathbf{1}}, \overline{\mathbf{1'}}} \end{array} &
\begin{array}{lll} \mathbf{1} & \mathbf{1'} & \mathbf{1''} \\ \overline{\mathbf{1}} & \multicolumn{2}{l}{r \nsupseteq \overline{\mathbf{1'}}, \overline{\mathbf{1''}}} \end{array} &
\begin{array}{lll} \mathbf{1} & \mathbf{2} &   \\ \overline{\mathbf{1}} & \multicolumn{2}{l}{r \neq \overline{\mathbf{2}}} \end{array}
\end{array}&
\begin{array}{lll} \mathbf{1} & \mathbf{1} & \overline{\mathbf{1}} \\ \overline{\mathbf{1}} & \multicolumn{2}{l}{r \nsupseteq \mathbf{1},\overline{\mathbf{1}}} \end{array} &
H^E_{12} V_{23} D^{-1}_{12} &
\text{none} \, (\mathbf{13})
\\
\hline
\begin{array}{c} (\,\makebox[\myem]{$\epsilon A$}\,\makebox[\myem]{$0$}\makebox[\myl]{$0$}) \\ (0\;\,\makebox[\myl]{$a$}\;a) \end{array} &
\text{IH} &
\hspace{-1.5mm}\begin{array}{lll}
\begin{array}{lll} \mathbf{1} & \mathbf{1} & \overline{\mathbf{1}} \\ \multicolumn{3}{l}{r \nsupseteq \mathbf{1}, \overline{\mathbf{1}}} \end{array} &
\begin{array}{lll} \mathbf{1'} & \mathbf{1} & \overline{\mathbf{1}} \\ \multicolumn{3}{l}{r \nsupseteq \mathbf{1}, \overline{\mathbf{1}}, \overline{\mathbf{1'}}} \end{array} &
\begin{array}{lll} \mathbf{1} & 2 &   \\ \multicolumn{3}{l}{r \nsupseteq \overline{\mathbf{1}}, 2} \end{array}
\end{array} &
\begin{array}{lll} \mathbf{1} & \mathbf{1} & \overline{\mathbf{1}} \\ \overline{\mathbf{1}} & \multicolumn{2}{l}{r \nsupseteq \mathbf{1},\overline{\mathbf{1}}} \end{array} &
H^E_{12} V_{23} D^{-1}_{12} &
\text{none} \, (\mathbf{13})
\\
\hline
\begin{array}{c} (\makebox[\myem]{$A$}\;\,\makebox[\myem]{$\epsilon B$}\;\,\makebox[\myl]{$0$}) \\ (\makebox[\myem]{$\epsilon a$}\;\makebox[\myem]{$\epsilon b$}\;\makebox[\myl]{$\epsilon c$}) \end{array} &
\text{NH or IH} &
\hspace{-1.5mm}\begin{array}{lll}
\begin{array}{lll} \mathbf{1} & \mathbf{1} & \mathbf{1} \\ \overline{\mathbf{1}} & \multicolumn{2}{l}{r \nsupseteq \overline{\mathbf{1}}} \end{array} &
\begin{array}{lll} \mathbf{1} & \mathbf{1} & \mathbf{1'} \\ \overline{\mathbf{1}} & \multicolumn{2}{l}{r \nsupseteq \overline{\mathbf{1}}, \overline{\mathbf{1'}}} \end{array}  &
\begin{array}{lll} \mathbf{1'} & \mathbf{1} & \mathbf{1} \\ \overline{\mathbf{1'}} & \multicolumn{2}{l}{r \nsupseteq \overline{\mathbf{1}}} \end{array} \vspace{1.2pt}  \\
\begin{array}{lll} \mathbf{1} & \mathbf{1'} & \mathbf{1''} \\ \overline{\mathbf{1}} & \multicolumn{2}{l}{r \nsupseteq \overline{\mathbf{1'}}, \overline{\mathbf{1''}}} \end{array}  &
\begin{array}{lll} \mathbf{1} & \mathbf{2} &   \\ \overline{\mathbf{1}} & \multicolumn{2}{l}{r \neq \overline{\mathbf{2}}} \end{array}  &
\end{array}&
\begin{array}{lll} 1 & 1 & 1 \\ 1 & 1 & r \neq 1 \end{array} &
V &
\text{none}
\\
\hline
\begin{array}{c} (\makebox[\myem]{$A$}\;\,\makebox[\myem]{$\epsilon B$}\;\,\makebox[\myl]{$0$})  \\ (0\;\makebox[\myem]{$\epsilon a$}\;\,\makebox[\myem]{$\epsilon a$}) \end{array} &
\text{IH} &
\hspace{-1.5mm}\begin{array}{lll}
\begin{array}{lll} \mathbf{1} & \mathbf{1} & \mathbf{1'} \\ \overline{\mathbf{1}} & \multicolumn{2}{l}{r \nsupseteq \overline{\mathbf{1}}, \overline{\mathbf{1'}}} \end{array} &
\begin{array}{lll} \mathbf{1} & \mathbf{1'} & \mathbf{1''} \\ \overline{\mathbf{1}} & \multicolumn{2}{l}{r \nsupseteq \overline{\mathbf{1'}}, \overline{\mathbf{1''}}} \end{array} &
\begin{array}{lll} \mathbf{1} & \mathbf{2} &   \\ \overline{\mathbf{1}} & \multicolumn{2}{l}{r \neq \overline{\mathbf{2}}} \end{array}
\end{array}&
\begin{array}{lll} \mathbf{1} & \mathbf{1} & \overline{\mathbf{1}} \\ \overline{\mathbf{1}} & \overline{\mathbf{1}} & r \neq \mathbf{1} \end{array} &
V_{23} D^{-1}_{12} &
\mathbf{13}
\\
\hline
\end{array}
\]
\caption{Lepton mass patterns that can be obtained starting from a symmetric limit ($\epsilon = 0$) in which either the neutrino or the charged lepton masses (but not both) vanish. The corrections proportional to $\epsilon$ are induced by the spontaneous symmetry breaking $G\to H$, under the hypothesis introduced in section~\ref{sec:breaking}. The corresponding irrep decompositions of $\rep^G_l$, $\rep^G_{e^c}$ and of $\rep^H_l$, $\rep^H_{e^c}$ leading to a viable form of the PMNS matrix are also shown. As usual, boldface fonts denote complex or pseudoreal (if 2-dimensional) irreps, primes are used to distinguish inequivalent representations, and in the case of complex representations $\mathbf{1}'$ is supposed to be different from both $\mathbf{1}$ and $\overline{\mathbf{1}}$. The representations of $G$ and $H$ are of course different even if represented by the same symbol. If $\epsilon$ is reabsorbed into the parameter it multiplies, the mass pattern correspond to the ones in the first four lines of \tab{PMNS} and the irrep decompositions of $\rep^H_l$, $\rep^H_{e^c}$ coincide with those shown in that table.}
\label{tab:GtoH}
\end{table}

\section{Constraints from unification}
\label{sec:unification}

A theory of flavour should account for both lepton and quark masses. The results we obtained provide constraints on the flavour group following from the observed pattern of lepton masses and mixings. The quark sector can of course provide additional constraints. 

In the context of unified theories, the two problems cannot be considered separately, as quarks and leptons are unified in single irreps of the unified gauge group. For example, in SU(5) theories, the lepton doublets $l_i$ are unified with the down quark singlets $d^c_i$ in anti-fundamental representations of SU(5), and the remaining fermions are unified into antisymmetric representations of SU(5). If the action of the flavour group commutes with SU(5), all fermions in the same SU(5) irrep should transform in the same way under the flavour group, $\rep_{d^c} = \rep_l$ and $\rep_{u^c} = \rep_q = \rep_{e^c}$. This provides an unavoidable further constraint on the flavour group and its representation. The constraint is even stronger if all the fermions of a single family are unified into a spinorial representation of SO(10). In this section we discuss the effect of such constraints on the previous results. 

Let us first assume that the flavour group commutes with SU(5) and call $\rep_{\overline{5}}$, $\rep_{10}$ its representations on the SU(5) fermion multiplets. As we have seen, the requirement that the prediction for lepton masses and mixings in the symmetric limit is close to what observed restricts the possible choices of $\rep_{\overline{5}} = \rep_l$ and $\rep_{10} = \rep_{e^c}$. \Tab{PMNS} summarises the 6 possible forms of their decompositions. Let us now require that the quark masses and mixings are also close what observed in the symmetric limit. By that we mean a quark mass pattern in the form $(A,0,0)$ or $(A,B,0)$ or $(A,B,C)$ in both the up and down quark sector, with the $(A,0,0)$ pattern preferred, as the others require hierarchies among the non-vanishing entries. As for the CKM matrix, let us first remind that the CKM angles are all measured to be small, with the only possible exception of the Cabibbo angle, corresponding to the 12 block of the CKM matrix. We then only consider the cases leading to a CKM matrix which is either diagonal or containing at most a non-trivial 12 block in the symmetric limit. It turns out that the only possible irrep decomposition is $\rep_{\bar{5}} = 1 + 1 + 1$,  $\rep_{10} = 1 + r \nsupseteq 1$. This uniquely identifies the form of the lepton spectrum in the symmetric limit, with vanishing electron and muon masses, $(A,0,0)$, and anarchical neutrino masses, $(a,b,c)$, with a generic PMNS matrix. The structure of the quark masses and mixings in the symmetric limit instead depends on the specific choice of $\rep_{10}$. This is shown in \tab{CKM1}, where the viable forms of $\rep_{10}$ and the corresponding mass and mixing patterns are listed. The down quark masses are in the same form as, and are actually equal to the charged lepton ones in the symmetric limit, as dictated by SU(5). The CKM matrix has the form 
\begin{equation}
\V = H^{\phantom{\dagger}}_U P^{\phantom{\dagger}}_U V P_D^{-1} H_D^{-1} \, .
\label{eq:mostgeneralV}
\end{equation}
The contributions to $\V$ have similar origins as the corresponding ones in \eq{mostgeneral}. As in the case of the PMNS matrix, each of them can be obtained without the need of writing explicitly nor diagonalising the quark mass matrices, with analogous rules. The form of the CKM matrix in terms of those contributions is also indicated in \tab{CKM1}. Note the constant presence of an undetermined transformation in the 12 block, $H^D_{12}$, associated to the vanishing of the two lighter down quark masses in the symmetric limit. As discussed, such undetermined transformations are fixed, up to diagonal phases, by symmetry breaking effects, and they can end up contributing to the Cabibbo angle with a zero, small, or large mixing angle. The patterns shown in the table are viable provided that the permutations $P_U$, $P_D$ do not modify the position of the heavy eigenvalue. The Cabibbo angle is expected to be large (with the measured value accidentally smallish) in the last case in \tab{CKM1}, where a physical $V_{12}$ rotation appears, which will survive symmetry breaking. In all the other cases, the Cabibbo angle can end up being large or small, depending on the symmetry breaking effects. If the two light eigenvalues are permuted, the Cabibbo angle receives a $\pi/2$ contribution, which needs to be (partially) cancelled by other contributions. 

\begin{table}[h]
\[
\begin{array}{|l|c|c|c|}
\hline
\begin{array}{lll} (\rep_{\bar{5}} \, = 1 & 1 & 1) \end{array} &
\text{masses}  & \V & \U  \\
\hline
\begin{array}{lll} \rep_{10} = 1 & \mathbf{1} & \mathbf{1} \\ \rep_{10} = 1 & \mathbf{1} & \mathbf{1'} \\ \rep_{10} = 1 & \mathbf{2} &  \end{array} &
\begin{array}{c} (\makebox[\myem]{$A$}\makebox[\myem]{$0$}\makebox[\myem]{$0$})_D \\ (\makebox[\myem]{$D$}\makebox[\myem]{$0$}\makebox[\myem]{$0$})_U \end{array} 
\, \begin{array}{c} (\makebox[\myem]{$A$}\makebox[\myem]{$0$}\makebox[\myem]{$0$})_E \\ (\makebox[\myem]{$a$}\makebox[\myem]{$b$}\makebox[\myem]{$c$})_\nu \end{array}
&
 H^U_{12} {H^D_{12}}^{-1}
& V
\\
\hline
\begin{array}{lll} \rep_{10} = 1 & 1' & \mathbf{1}  \end{array} &
\begin{array}{c} (\makebox[\myem]{$A$}\makebox[\myem]{$0$}\makebox[\myem]{$0$})_D \\ (\makebox[\myem]{$D$}\makebox[\myem]{$E$}\makebox[\myem]{$0$})_U \end{array} 
\, \begin{array}{c} (\makebox[\myem]{$A$}\makebox[\myem]{$0$}\makebox[\myem]{$0$})_E \\ (\makebox[\myem]{$a$}\makebox[\myem]{$b$}\makebox[\myem]{$c$})_\nu \end{array}
&
 P^U_{2\leftrightarrow 3} {H^D_{12}}^{-1}
& V
\\
\hline
\begin{array}{lll} \rep_{10} = 1 & \mathbf{1} & \overline{\mathbf{1}}  \end{array} &
\begin{array}{c} (\makebox[\myem]{$A$}\makebox[\myem]{$0$}\makebox[\myem]{$0$})_D \\ (\makebox[\myem]{$D$}\makebox[\myem]{$E$}\makebox[\myem]{$F$})_U \end{array} 
\, \begin{array}{c} (\makebox[\myem]{$A$}\makebox[\myem]{$0$}\makebox[\myem]{$0$})_E \\ (\makebox[\myem]{$a$}\makebox[\myem]{$b$}\makebox[\myem]{$c$})_\nu \end{array}
&
 P_U {H^D_{12}}^{-1}
& V
\\
\hline
\begin{array}{lll} \rep_{10} = 1 & 1' & 1''  \end{array} &
\begin{array}{c} (\makebox[\myem]{$A$}\makebox[\myem]{$0$}\makebox[\myem]{$0$})_D \\ (\makebox[\myem]{$D$}\makebox[\myem]{$E$}\makebox[\myem]{$F$})_U \end{array} 
\, \begin{array}{c} (\makebox[\myem]{$A$}\makebox[\myem]{$0$}\makebox[\myem]{$0$})_E \\ (\makebox[\myem]{$a$}\makebox[\myem]{$b$}\makebox[\myem]{$c$})_\nu \end{array}
&
 P_U {H^D_{12}}^{-1}
& V
\\
\hline
\begin{array}{lll} \rep_{10} = 1 & 1' & 1'  \end{array} &
\begin{array}{c} (\makebox[\myem]{$A$}\makebox[\myem]{$0$}\makebox[\myem]{$0$})_D \\ (\makebox[\myem]{$D$}\makebox[\myem]{$E$}\makebox[\myem]{$F$})_U \end{array}  
\, \begin{array}{c} (\makebox[\myem]{$A$}\makebox[\myem]{$0$}\makebox[\myem]{$0$})_E \\ (\makebox[\myem]{$a$}\makebox[\myem]{$b$}\makebox[\myem]{$c$})_\nu \end{array}
&
 P_U
 V_{12}
{H^D_{12}}^{-1}
& V
\\
\hline
\end{array}
\]
\caption{Possible forms of SU(5) unified flavour representations. $\rep_{\overline{5}}$ is trivial in all cases. The form of fermion masses and of the CKM and PMNS matrices, in the notations of \eq{mostgeneralV}, corresponding to viable choices are shown. The lepton mass pattern and PMNS matrix are all in the same form, as they all correspond to the case in the first line of \tab{PMNS}. $P_{2 \leftrightarrow 3}$ is either the identity permutation or the switch of $2$ and $3$. }
\label{tab:CKM1}
\end{table}

If all the fermions of a single family are unified into a dimension 16 spinorial representation of SO(10) commuting with the flavour group, the constraints on the flavour group representation are even stronger, and no solution can be found. In such a case we would have in fact $\rep_{16} \equiv \rep_{\overline{5}} = \rep_{10}$. The symmetric limit is a good approximation in the lepton sector only if $\rep_{16}$ is trivial. Such a possibility however leads to a generic CKM matrix with $\ord{1}$ angles, which we do not consider a viable leading order approximation in the symmetric limit.

\section{Conclusions}
\label{sec:conclusions}

We provided a complete answer to the following general question: what are the flavour groups, of any type, and representations providing, in the symmetric limit, an approximate description of lepton (fermion) masses and mixings? 

The assumption we made is quite general: the light neutrinos are of Majorana type, and the symmetry arguments can be applied directly to their mass matrix. Despite the generality of the problem, the complete answer is simple and has an important corollary: either the flavour symmetry does not constrain at all the neutrino mass matrix (anarchy), or the neutrinos have an inverted hierarchical spectrum. Therefore, if the present hint of a normal hierarchical spectrum were confirmed, we would conclude that, under the above assumption, flavour models leading to an approximate description of lepton masses and mixings in the symmetric limit are not able to account for any of the neutrino flavour observables, and symmetry breaking effects must play a primary role in their understanding. Such a conclusion is further strengthened in the case in which the representation of the flavour group commutes with the standard representation of a SU(5) grand unified gauge group. In the latter case, not even the options leading to an inverted hierarchical spectrum are available, and the only option is anarchy. In the case of SO(10), there are no solutions. 

The main caveat to the previous conclusion is the assumption that the light neutrinos are of Majorana type, and that the symmetry arguments can be applied directly to their mass matrix. The origin of Majorana neutrino masses most likely resides at high scales, where additional relevant degrees of freedom (singlet neutrinos for example) might live. In such a case, the flavour symmetry acts on the high-scale degrees of freedom as well. The low-energy analysis turns out to be often equivalent to the high-scale analysis, but not always. Such a caveat will be studied in future work. 

The possibility to provide a simple systematic answer to the above general question is based on the following result: the structure of lepton masses and mixings only depends on the flavour group and representations through the structure of their decomposition in irreducible components, and in particular only through the dimension, type (complex or real or pseudoreal), and equivalence of those components. We found that there are only six viable structures, listed in \tab{PMNS}. All of them contain only one-dimensional real or complex representations. 

In passing, we developed a simple technique to determine the form of the lepton masses and mixings directly from the structure of the decomposition in irreducible representations, without the need to specify, nor to diagonalise, the lepton mass matrices. We also noted that it is important to write the invariance condition in terms of the charged lepton mass matrix $m_E$ and not of $m^\dagger_E m^{\phantom{\dagger}}_E$, otherwise the important role of the flavour representation on singlet leptons would be lost. 

As our results and assumptions imply that an understanding of the flavour observables of normal hierarchical neutrinos must rely on symmetry breaking effects, we also consider the possibility that the neutrino or the charged lepton mass matrix vanishes in the symmetric limit. With a simple extension of the previous techniques, we proved that the sole knowledge of the symmetry breaking pattern, i.e.\ of the residual unbroken group, is not sufficient to get a better understanding of the flavour observables: the sources of flavour breaking and of their vacuum expectation values need to be specified.

\section*{Acknowledgments}

We thank Ferruccio Feruglio and Alexander Stuart for interesting and useful discussions. The work of A.R.\ was partially supported by the PRIN project 2015P5SBHT\_002 and by the EU Horizon 2020 Marie Sk\l odowska-Curie grant agreements No.\ 690575 and 674896. 

\appendix

\addtocontents{toc}{\protect\setcounter{tocdepth}{1}}

\section{Proof of the results in section~\ref{sec:patterns}}
\label{sec:proofs}

In this appendix, we find the general form of the PMNS matrix associated to a generic decomposition of $\rep_l$ and $\rep_{e^c}$ in irreducible components. We consider the general case of $n$ families. 

Let us first introduce a few notations. The irreducible components of $\rep_l$ are of different, possible inequivalent types. A given irrep type ``$r$'', of dimension $d_r$, can appear in the decomposition of $\rep_l$ more than once. We denote with $n_r$ the number of times it appears. Analogously, $n^c_r$ is the number of times the irrep $r$ appears in the decomposition of $\rep_{e^c}$. 

Given a lepton doublet $l_i$, we can then associate three labels to it. We can denote by $r$ the type of irrep to which $l_i$ belongs. As each type of representation may appear more than once in the decomposition of $\rep_l$, we can denote by  $k$  the occurrence to which $l_i$ belongs ($1\leq k \leq n_r$). Finally, as the irrep $r$ may have dimension larger than 1, we can denote by $a$ the position of the lepton $l_i$ within its irrep multiplet ($1\leq a \leq d_{r}$). All in all, the lepton $l_i$ is identified by its ``irrep coordinates'' $(r,k, a)$. Such  coordinates can be used as an alternative labelling of the lepton doublets $l_i$ (and of its components $e_i,\nu_i$). The generic lepton doublet will in this case be denoted by $l_{rka}$. Clearly, there is a correspondence between the two possible labelling, the one by $1\leq i \leq n$ and the one by $rka$, defined by 
\begin{equation}
l_i = l_{rka} \; .
\label{eq:irrepcoor}
\end{equation}

Analogous coordinates $(r,k,a)$ can be used to identify the lepton singlets $e^c_i$. The irreps $r$ found in the decomposition of $\rep_{e^c}$ can be different than the ones found in $\rep_l$, and their multiplicities in the decompositions can also be different. 

We can, and will, choose a flavour basis for the leptons $l_i$ and $e^c_i$, and the mappings between the ``$i$''  and the ``$(rka)$'' indices, as follows. 
\begin{itemize}
\item
Each irrep of type $r$ acts on a set of subsequent leptons $(l_{i_0}\ldots l_{i_0+d_r})$, forming a certain occurrence $k_0$ of the irrep type $r$, $(l_{i_0}\ldots l_{i_0+d_r}) = (l_{rk_01} \ldots l_{rk_0d_r})$. 
\item
As stated in section~\ref{sec:masses}, non-vanishing charged lepton masses correspond to conjugated irreps in the decompositions of $\rep_l$ and $\rep_{e^c}$. Consider then the copies $k =1\ldots n_r$ of a certain irrep type $r$ in $\rep_l$ and the copies $h=1\ldots n^c_{\bar r}$ of the conjugated representation $\bar r$ in $\rep_{e^c}$ ($\bar r = r$ if $r$ is real or pseudoreal). Only a number $\min(n_r,n^c_{\bar r})$ of them can be paired to get possibly non-vanishing masses, while all the residual unpaired leptons are forced to be massless. We assume that the leptons $l_{rka}$ and $e^c_{\bar r ka}$ occupy the same positions in the lists $l_1\ldots l_n$ and $e^c_1\ldots e^c_n$, for all $k \leq \min(n_r,n^c_{\bar r})$. Tables~\ref{tab:irreppattern1}, \ref{tab:irreppattern2} use such a convention. 
\item
All irreps of type $r$ are represented by the same $d_r\times d_r$ unitary matrix $U_r$ on the corresponding leptons: $l_{rka} \to U^r_{ab} l_{rkb}$, $e^c_{rka} \to (U^{\bar r}_{ab})^* e^c_{rkb}$.\footnote{In practice: if $r$ is real or complex, it has the same action on lepton doublets and singlets, as $(U^{\bar r}_{ab})^* = U^r_{ab}$;  if $r$ is pseudoreal, it acts  on the singlets in the conjugated (but equivalent) way, as $(U^{\bar r}_{ab})^* = (U^r_{ab})^*$.} If $r$ is real, the matrix $U$ is real; if $r$ is complex, $U_{\bar r} = (U_r)^*$; if $r$ is pseudoreal, $\omega \, U_r = U^*_r \, \omega$, where
\begin{equation}
\setlength\arraycolsep{4pt}
\omega = 
\begin{pmatrix}
0 & 1 & & & \\
-1 & 0 & & & \\
& & \ddots & & \\
& & & 0 & 1 \\
& & & -1 & 0
\end{pmatrix},
\label{eq:omega}
\end{equation}
is a $d_r \times d_r$ antisymmetric block matrix and $d_r$ is even for pseudoreal representations. \end{itemize}

Having set up the necessary notations, we are now ready to discuss the structure of the lepton mass matrices in the above basis. A non-zero entry $m^E_{ij} \neq 0$ paring the leptons $e^c_i$ and $e_j$  is allowed only when the irrep to which $e^c_i$ and $l_j$ belong are conjugated, say $r$ and $\overline{r}$ respectively. If $r$ or $\bar r$ appear more than once in the decomposition of $\rep_l$ or $\rep_{e^c}$, the non-zero entries form a rectangular block, of size $n^c_{\bar r} \times n_r$, whose entries we can denote by $m^{E,r}_{kh}$. If the irrep $r$ has dimension $d_r > 1$, $m^{E,r}_{kh}$ is the common diagonal element for all the leptons in the corresponding multiplet. Such a structure becomes transparent when the mass matrices are written in terms of the irrep coordinates. Indeed, the invariance under $G$ forces the charged lepton mass matrix to be in the form
\begin{equation}
m^E_{rka,shb} =\delta_{\overline{r} s} \delta_{ab} \, m^{E,r}_{k h}. 
\label{eq:mErho}
\end{equation}
Conversely, any charged lepton mass matrix in that form is of course invariant. Analogously, the form of the neutrino mass matrix is
\begin{equation}
m^\nu_{rka,shb} =
\left \{
\begin{aligned}
& \delta_{\overline rs} \delta_{ab} m^{\nu,r}_{kh} &&\text{if $r,s$ both complex} &&\text{($m^{\nu,r}$ generic)} \\
& \delta_{rs} \delta_{ab} m^{\nu,r}_{kh} &&\text{if $r,s$ both real} &&\text{($m^{\nu,r}$ symmetric)} \\
& \delta_{rs} \omega_{ab} m^{\nu,r}_{kh} &&\text{if $r,s$ both pseudoreal} &&\text{($m^{\nu,r}$ antisymmetric)} \\
& 0 &&\text{if $r,s$ of different type} &&
\end{aligned}
\right .
\label{eq:mnurho}
\end{equation}

Note that the entries $m^{\nu,r}_{kh}$ appear in off diagonal positions, unless the representation $r$ is real. This is of course because of the Majorana nature of the neutrino mass matrix. Diagonal entries are allowed in the symmetric limit only when the representation to which the corresponding lepton belongs is real. 

Note also that pseudoreal representations are only marginally relevant in the three neutrino case. As the dimension of pseudoreal representations is even, there is room for at most one pseudoreal irrep in that case. Moreover, if one two-dimensional pseudoreal representation appears in $\rep_l$, the two rows and columns of the neutrino mass matrix corresponding to that representation vanish, as $m^{\nu,r}$ in \eq{mnurho} is a $1\times 1$ antisymmetric matrix, so that $m^{\nu,r} = 0$. Still, we will stick in the following for completeness to the $n$ neutrino case and to the full treatment of the pseudoreal case. 

The PMNS matrix arises from the diagonalisation of  $m^E_{ij}$ and $m^\nu_{ij}$ in eqs.~(\ref{eq:mErho},\ref{eq:mnurho}). It is made of four types of contributions, each with a different physical origin:
\begin{enumerate}
\item 
A core contribution $V$ associated to the presence of equivalent irreps in the lepton doublet representation $\rep_l$.
\item
A contribution $D$ associated to the possible presence of Dirac structures in $m^\nu$ and providing maximal mixing. 
\item
Permutations $P$ associated to the requirement that charged lepton and neutrino masses need to be in a standard ordering. 
\item
``Unphysical'' contributions $H$ associated to the arbitrariness in the choice of the basis in flavour space for degenerate leptons. 
\end{enumerate}
Let us see how such contributions arise from the diagonalisation of  $m_E$ and $m_\nu$.

\subsection{V}
\label{sec:V}

The first contribution $V$ to the PMNS matrix is a unitary matrix commuting with $\rep_l$. Such a unitary matrix $V$ mixes lepton multiplets belonging to identical irreps and is non-trivial only if the decomposition $\rep_l$ contains more than one copy of the same irrep. All possible forms of $V$ compatible with the previous requirements can be obtained. 

In order to show how $V$ arises, we observe that $m_\nu$, $m_E$ can be diagonalised, up to Dirac structures in the neutrino sector (we will see below what this means) by unitary transformations of the charged leptons and neutrinos $\nu_i$, $e_i$, $e^c_i$ commuting with the action of $G$,
\begin{equation}
\begin{aligned}
\nu'_{rka} &= V^{\nu,r}_{kh} \nu_{rha} \\
e'_{rka} &= V^{e,r}_{kh} e_{rha} \\
e^{c\,\prime}_{rka} &= V^{e^c,r}_{kh} e^{c}_{rha}.
\end{aligned}
\label{eq:Vblock}
\end{equation}
$V^{\nu,r}$, $V^{e,r}$, $V^{e^c,r}$ are squared matrices and $V^{\nu,r}$, $V^{e,r}$ have the same dimension $n_r$. They mix full equivalent multiplets (they do not act on the index $a$) and are non-trivial in the presence of more than one copy of the representation $r$. The above transformations can be chosen to diagonalise each of the blocks in eqs.~(\ref{eq:mErho},\ref{eq:mnurho}) as follows. 

In the case of charged lepton blocks, we have
\begin{equation}
m_{E,r} = V_{e^c,\bar r}^T \, m_{E,r}^\text{diag} \, V^{\phantom{T}}_{e,r} 
\; .
\label{eq:blockE}
\end{equation}
As for the neutrino blocks, we need to treat the pseudoreal case differently. In the case of real or complex representations, we have
\begin{equation}
m_{\nu,r} = V_{\nu,\bar r}^T \, m_{\nu,r}^\text{diag} \, V^{\phantom{T}}_{\nu,r}
\; .
\label{eq:blocknu}
\end{equation}
If $r$ is real, $m_{\nu,r}$ is a symmetric complex matrix, and \eq{blocknu} gives its diagonalisation in terms of a single unitary transformation $V_{\nu,r}$. If $r$ is complex, the block is in general rectangular, $m^{\phantom{T}}_{\nu,\bar r} = m_{\nu,r}^T$, and \eq{blocknu} gives the diagonalisation of both in terms of two independent complex matrices $V_{\nu,r}$ and $V_{\nu,\bar r}$ of dimension $n_r$ and $n_{\bar r}$ respectively. When the matrices $m_{E,r}^\text{diag},  m_{\nu,r}^\text{diag}$ above are rectangular, we conventionally choose the non-vanishing eigenvalues to appear on the diagonal starting from the lower-right corner. For example, if there are more columns than rows
\begin{equation*}
m^\text{diag} = \begin{pmatrix}
0 & \cdots & 0 & X & & 0  \\
& & & & \ddots \\
0 & \cdots & 0 & 0 & & X 
\end{pmatrix}\;,
\end{equation*}
where $X$ denotes the position of the eigenvalues. \Eqs{blockE} and~(\ref{eq:blocknu}) define $V_{e,r}$ ($V_{\nu,r}$) for each irrep type $r$ found in the decomposition of $\rep_l$, provided that $\bar r$ is also found in the decomposition of $\rep_{e^c}$ ($\rep_l$), so that the block to be diagonalised exists. If this is not the case, we define $V_{e,r}$ ($V_{\nu,r}$) to be the identity matrix.

Let us now consider the special case of a neutrino block corresponding to a pseudoreal representation $r$. In such a case, $m_{\nu,r}$ is a square, $n_r \times n_r$ antisymmetric matrix. It can be reduced to the following ``pseudo-diagonal'' form
\begin{equation}
m_{\nu,r} = V^T_{\nu,r} \, m_{\nu,r}^\text{ps-diag} \, V^{\phantom{T}}_{\nu,r} \;,
\quad
(m_{\nu,r}^\text{ps-diag})_{kh} = m^{\nu,r}_k \omega_{kh} \;,
\quad
m^{\nu,r}_{2\kappa} = m^{\nu,r}_{2\kappa - 1} \geq 0 \;.
\label{eq:blockps}
\end{equation}
The matrix $\omega$ can now have even or odd dimension, depending on the number of copies $n_r$ of the irrep $r$. If $n_r$ is odd, $\omega$ is the restriction to the first $n_r$ rows and columns of a matrix $\omega$ of larger even dimension, which means that it is in the form in \eq{omega}, with the addition of one extra vanishing row and column. The matrix $m_{\nu,r}^\text{ps-diag}$ is therefore an antisymmetric block diagonal matrix, with subsequent $2\times 2$ blocks in the form 
\[
\begin{pmatrix}
0 & m^{\nu,r}_k \\-m^{\nu,r}_k & 0
\end{pmatrix} \;,
\]
possibly followed by a singly vanishing diagonal entry if $n_r$ is odd. Therefore, the pseudoreal irreps are now paired in couples $(12)$, $(34)$, \ldots, $(2\kappa-1,2\kappa)$, \ldots, each associated to degenerate masses, with a possibly unpaired last irrep (if the total number is odd) associated to a zero mass. 

All in all, we have
\begin{equation}
m_E = V_{e^c}^T \, m_E^\text{diag} \, V^{\phantom{T}}_e ,
\qquad
m_\nu = V_\nu^T \, m_\nu^\text{s-diag} \, V^{\phantom{T}}_\nu ,
\label{eq:blockdiag2}
\end{equation}
where
\begin{equation}
V^\nu_{ij} = \delta_{\bar r s} \delta_{ab} V^{\nu,s}_{kh} 
\qquad
V^e_{ij} = \delta_{\bar rs} \delta_{ab} V^{e,s}_{kh}
\qquad
V^{e^c}_{ij} = \delta_{\bar rs} \delta_{ab} V^{e^c,s}_{kh} , \\
\label{eq:mdiagfull}
\end{equation}
and $i\leftrightarrow (rka)$ and $j\leftrightarrow (shb)$, as defined by \eq{irrepcoor}. Clearly, $V_e$ and $V_\nu$ commute with $\rep_l$. We can now define 
\begin{equation}
V = V^{\phantom{\dagger}}_e V^\dagger_\nu \;,
\label{eq:V}
\end{equation}
which represents the core contribution to the PMNS matrix and also commutes with $\rep_l$. 

\Eq{blockdiag2} brings the charged lepton mass matrix in diagonal form,
\begin{equation}
(m^\text{diag}_E)_{ij} = \delta_{\overline{r}s} \delta_{kh} \delta_{ab}m^{E,r}_{h}  \;.
\label{eq:mdiagfullE}
\end{equation}
The eigenvalues do lie on the diagonal because of the assumptions we made on the ordering of the charged leptons. The leptons $e'_{rka}$ get mass $m^{E,r}_{k}$ by pairing with $e^{c\,\prime}_{\bar rka}$. If the multiplet has dimension $d_r >1$, all the leptons in the multiplets end up being degenerate. As the number of representations of type $r$ acting on the lepton doublets, labelled by $k=1\ldots n_r$, and the number of representations of type $\bar r$ acting on the lepton singlets, labelled by $k=1\ldots n^c_{\bar r}$, can be different, only the first $k=1\ldots \min(n_r,n^c_r)$ pairs get a possibly non-zero mass, while all residual unpaired charged leptons are forced to be massless. 

\Eq{blockdiag2} brings the neutrino mass matrix in a ``semi-diagonal'' form, 
\begin{equation}
(m^\text{s-diag}_\nu)_{ij}  = 
\left\{
\begin{aligned}
&\delta_{\overline{r}s} \delta_{kh} \delta_{ab} m^{\nu,r}_{k} && \text{if neither $r$ nor $s$ is pseudoreal} \\
&\delta_{\overline{r}s} \omega_{kh} \omega_{ab} m^{\nu,r}_{k} && \text{if both $r$ and $s$ are 
pseudoreal} \\
&0 && \text{otherwise}
\end{aligned}
\right. ,
\label{eq:mdiagfullnu}
\end{equation}
where again $m^{\nu,r}_{2\kappa -1} = m^{\nu,r}_{2\kappa}$ in the pseudoreal case ($\kappa$ integer). 

All neutrinos $\nu'_{rka}$ corresponding to real representations $r$ get a diagonal (Majorana) mass term $m^{\nu,r}_{k}$ by pairing to themselves. If the representation has dimension $d_r > 1$, all neutrinos in the multiplets are degenerate. The neutrinos $\nu'_{rka}$ corresponding to complex representations $r$ get a Dirac mass term $m^{\nu,r}_{k} = m^{\nu,\bar r}_{k}$ by pairing to the neutrinos in $\nu'_{\overline{r} ka}$ in the conjugated representation $\bar rk$. If $d_r >1$, all the neutrinos in the two conjugated multiplets are degenerate. As the number of representations of type $r$, labelled by $k=1\ldots n_r$, and the number of representations of type $\bar r$, labelled by $k=1\ldots n_{\bar r}$, can be different, only the first $k=1\ldots \min(n_r,n_{\bar r})$ pairs get a possibly non-zero mass, while all residual unpaired neutrinos are forced to be massless. Finally, in the case of pseudoreal representations, the two pairs of neutrinos $\nu'_{r, 2\kappa, 2\alpha}$, $\nu'_{r, 2\kappa-1, 2\alpha-1}$ and $\nu'_{r, 2\kappa, 2\alpha-1}$, $\nu'_{r, 2\kappa-1, 2\alpha}$ both get a Dirac mass term, both with mass $m^{\nu,r}_{2\kappa} = m^{\nu,r}_{2\kappa-1}$. If $d_r >1$, all the neutrinos in the two paired multiplets $k = 2\kappa$ and $k = 2\kappa -1$ are degenerate. For $n_r$ odd, two spare neutrinos are massless. 

To summarise, $m_\nu^\text{s-diag}$ is not necessarily diagonal because of the possible presence of Dirac structures associated to paired conjugated and pseudoreal representations, and its non-vanishing entries can be found:
\begin{itemize}
\item
In all the diagonal positions $m^\text{s-diag}_{rka, rka}$ corresponding to real irreps $r$, providing a Majorana mass term for the neutrino $\nu'_{rka}$.
\item
In symmetric off-diagonal positions, $m^\text{s-diag}_{\bar r k a,r k a} = m^\text{s-diag}_{r k a,\bar r k a}$, corresponding to complex representations $r$ and $k \leq \min(n_r,n_{\bar r})$, providing a Dirac mass term to the conjugated neutrinos $\nu'_{rka}$ and $\nu'_{\bar r k a}$.
\item
In symmetric off-diagonal positions $m^\text{s-diag}_{r (2\kappa)(2 \alpha), r, (2\kappa-1) (2\alpha-1)} = m^\text{s-diag}_{r (2\kappa-1)(2 \alpha -1), r (2\kappa) (2\alpha)} = -m^\text{s-diag}_{r (2\kappa)(2 \alpha-1), r (2\kappa-1)(2\alpha)} = -m^\text{s-diag}_{r (2\kappa-1)(2 \alpha), r(2\kappa)(2\alpha-1)}$, corresponding to pseudoreal representations $r$ and $\kappa = 1\ldots \lfloor n_r/2 \rfloor $, $\alpha = 1\ldots d_r/2$.
\end{itemize}

\subsection{D}
\label{sec:D}

In order to complete the diagonalisation of the lepton mass matrices, we need to diagonalise the Dirac structures in $m_\nu^\text{s-diag}$. This is how the contribution $D$ to the PMNS matrix, containing a maximal mixing transformation for each Dirac structure, arises. 

As discussed in the previous subsection, the semi-diagonal matrix $m_\nu^\text{s-diag}$ contains a diagonal block corresponding to the neutrinos $\nu_{rka}$ in real irreps $r$; a $2\times 2$ Dirac block corresponding to neutrinos in paired conjugated complex representations $\nu_{rka}$ and $\nu_{\bar r ka}$, $k=1\ldots \min(n_r,n_{\bar r})$; a trivially diagonal vanishing block corresponding to neutrinos in unpaired complex representations $\nu_{rka}$, $k > \min(n_r,n_{\bar r})$; a trivially diagonal vanishing block corresponding to the neutrinos $\nu_{r n_r a}$ in the last copy of the pseudoreal irrep $r$, if $n_r$ is odd (and in particular if there is only one copy of $r$); if there are at least two copies of $r$, a $4\times 4$ Dirac block corresponding to the four neutrinos $\nu'_{r, 2\kappa-1, 2\alpha-1}$, $\nu'_{r, 2\kappa, 2\alpha}$, $\nu'_{r, 2\kappa-1, 2\alpha}$, $\nu'_{r, 2\kappa, 2\alpha-1}$. The matrix $m_\nu^\text{s-diag}$ can then be diagonalised by diagonalising the above Dirac blocks as follows. 

As seen, there are two types of Dirac blocks, associated to complex conjugated and to pseudoreal irreps respectively (only the former are relevant to the three neutrino case, as the latter arises only in the presence of at least four neutrinos). 

In the case of a Dirac block associated to the neutrinos $\nu_{rka}$ and $\nu_{\bar r ka}$ in conjugated complex irreps, and for $k=1\ldots \min(n_r,n_{\bar r})$, $a = 1\ldots d_r$, the block has the form
\begin{equation}
\setlength\arraycolsep{4pt}
\begin{pmatrix}
0 & m^{\nu,r}_{k} \\
m^{\nu,r}_{k} & 0
\end{pmatrix} \; ,
\label{eq:Dirac}
\end{equation}
where $m^{\nu,r}_{k} \geq 0$ ($m^{\nu,r}_{k} = m^{\nu,\bar r}_{k}$). Its diagonalisation is trivial
\begin{equation}
\setlength\arraycolsep{4pt}
\begin{pmatrix}
0 & m^{\nu,r}_{k} \\
m^{\nu,r}_{k} & 0
\end{pmatrix} = D_2^T 
\begin{pmatrix}
m^{\nu,r}_{k}  & 0 \\
0 & m^{\nu,r}_{k} 
\end{pmatrix}
D_2 \;, \qquad
D_2 = 
\frac{1}{\sqrt{2}}
\begin{pmatrix}
1 & 1 \\
-i & i
\end{pmatrix}. 
\label{eq:D2}
\end{equation}
The unitary matrix $D_2$ corresponds to a maximal rotation by an angle $\pi/4$, together with a phase redefinition by the imaginary unit $i$, needed to make the diagonal entries positive. Such a Majorana phase is physical, but it plays a negligible role in oscillation experiments. The matrix $D_2$ is defined up to a phase, meaning that we could have equivalently used the following form of $D_2$,
\begin{equation}
\frac{1}{\sqrt{2}}
\begin{pmatrix}
e^{i\theta} & e^{-i\theta} \\ \mp i e^{i\theta} & \pm i e^{-i\theta} 
\end{pmatrix} \; .
\label{eq:D2tilde}
\end{equation}
The phase $\theta$ corresponds to the freedom to perform a O(2) transformation on the two degenerate neutrino mass eigenstates, and can be reabsorbed in a phase redefinition of $V_\nu$. The sign is unphysical.

In the case of a Dirac block associated to the two paired pseudoreal irreps $2\kappa -1$ and $2\kappa$ ($\kappa = 1\ldots \lfloor n_r/2 \rfloor$) and involving the four neutrinos $\nu'_{r, 2\kappa-1, 2\alpha-1}$, $\nu'_{r, 2\kappa, 2\alpha}$, $\nu'_{r, 2\kappa-1, 2\alpha}$, $\nu'_{r, 2\kappa, 2\alpha-1}$ (rows and columns of the matrix below ordered accordingly), the block has the form
\begin{equation}
\setlength\arraycolsep{1pt}
\left(
\begin{array}{cc|cc}
0 & m^{\nu,r}_{2k} & & \\
m^{\nu,r}_{2k} & 0 & & \\ \hline
& & 0 & -m^{\nu,r}_{2k} \\
& & -m^{\nu,r}_{2k} & 0 
\end{array} 
\right) = 
\setlength\arraycolsep{8pt}
\renewcommand{\arraystretch}{2}
\left(
\begin{array}{c|c}
\raisebox{3pt}{\text{\large $D_2$}} & \\ \hline
& \raisebox{1pt}{\text{\large $iD_2$}} 
\end{array}
\right)^T
\renewcommand{\arraystretch}{1}
\setlength\arraycolsep{1pt}
\left(
\begin{array}{cc|cc}
m^{\nu,r}_{2k} & & & \\
& m^{\nu,r}_{2k} & & \\ \hline
& & m^{\nu,r}_{2k} & \\
& & & m^{\nu,r}_{2k}
\end{array} 
\right)
\setlength\arraycolsep{8pt}
\renewcommand{\arraystretch}{2}
\left(
\begin{array}{c|c}
\raisebox{3pt}{\text{\large $D_2$}} & \\ \hline
& \raisebox{1pt}{\text{\large $iD_2$}} 
\end{array}
\right) ,
\label{eq:Diracps}
\end{equation}
where $m^{\nu,r}_{2\kappa} \geq 0$ ($m^{\nu,r}_{2\kappa} = m^{\nu,r}_{2\kappa-1}$). 

Based on what above, we can define unitary matrix $D$ to be the product of the (commuting) $2\times 2$ transformations $D_2$ acting on neutrinos in paired complex or pseudoreal representations. The matrix $D$ will therefore be diagonal in the block corresponding to the neutrinos in real irreps and in the block corresponding to the neutrinos in unpaired complex or pseudoreal representations; it will contain an instance of the matrix $D_2$ in each $2\times 2$ block corresponding to neutrinos $\nu_{rka}$ and $\nu_{\bar r ka}$ in paired conjugated complex representations, $k=1\ldots \min(n_r,n_{\bar r})$; and it will contain an instance of $D_2$ and $iD_2$ in each pair of $2\times 2$ blocks corresponding to the neutrinos ($\nu'_{r, 2\kappa-1, 2\alpha-1}$, $\nu'_{r, 2\kappa, 2\alpha}$) and ($\nu'_{r, 2\kappa-1, 2\alpha}$, $\nu'_{r, 2\kappa, 2\alpha-1}$) respectively, in paired pseudoreal representations, $\kappa = 1\ldots \lfloor n_r/2 \rfloor$. 

As a consequence, the semi-diagonal matrix $m_\nu^\text{s-diag}$ is diagonalised as follows
\begin{equation}
m_\nu^\text{s-diag} = D^T m_\nu^\text{diag} D\;,
\label{eq:Ddiag}
\end{equation}
where $m_\nu^\text{diag}$ is diagonal, with degenerate eigenvalues in the positions corresponding to neutrinos in paired complex conjugated or pseudoreal representations. 

\subsection{P}
\label{sec:P}

What above provides a full diagonalisation of the lepton mass matrix in terms of the unitary transformations $V_e$, $V_{e^c}$ and $(D V_\nu)$:
\begin{equation}
m_E = V_{e^c}^T \, m_E^\text{diag} \, V^{\phantom{T}}_e ,
\qquad
m_\nu = (D V_\nu)^T \, m_\nu^\text{diag} \, (D V^{\phantom{T}}_\nu) .
\label{eq:fulldiag}
\end{equation}
We are therefore close to identifying the PMNS matrix. In order to do that, we should take into account the fact that the order of the rows and columns of the PMNS matrix is defined by a standard ordering of the leptons. In the case of charged leptons, the standard ordering coincides with the mass ordering, $m_{e_1} \leq \ldots \leq m_{e_n}$. In the three neutrino case, the standard ordering for neutrinos defines the mass eigenstates $\nu_1$ and $\nu_2$ to be the two ones closer in terms of squared mass difference, with $\nu_1$ being the lightest of the two. In order to find the PMNS matrix, we should then permute the lepton mass eigenstates in order to have them in the standard ordering. This is achieved by two permutation matrices $P_E$ and $P_\nu$,
\begin{equation}
m_E^\text{diag} = P_E^T m_E^\text{diag,so} P_E^{\phantom{T}}, 
\qquad
m_\nu^\text{diag} = P_\nu^T m_\nu^\text{diag,so} P_\nu^{\phantom{T}} ,
\label{eq:P}
\end{equation}
where ``so'' stands for ``standard ordering''. 

A few comments are in order. We are considering here the symmetric limit. On the other hand, the standard ordering is defined on the physical masses, which also get contributions from symmetry breaking effects. However, in the assumption we made that symmetry breaking effects are small, the ordering is not affected by symmetry breaking effects. 

An exception to the latter argument arises in the presence of degenerate eigenvalues (vanishing or not). Which linear combination of the corresponding leptons will end up being the lighter or heavier crucially depends in this case on the symmetry breaking effects. This type of ambiguity will be taken into account by the $H$ matrix defined in the next subsection, so that no permutation needs to be introduced. 

As an example in which a physical permutation is involved is when the charged lepton spectrum ends up being $(m_{e_3},m_{e_2},m_{e_1}) = (0,0,A)$ instead of $(m_{e_3},m_{e_2},m_{e_1}) = (A,0,0)$. In such a case a permutation $P^E_{1\to 3}$ moving the first lepton in the last position is necessary (such a permutation is defined up to a further permutation of the first two elements, but the latter does not need to be taken into account). In such a case, the permutation only depends on the mass pattern and not on the specific values of the non-zero entries. A physical permutation is also needed when the mass ordering depends on the specific values of the non-zero entries, for example if $(m_{e_3},m_{e_2},m_{e_1}) = (A,B,0)$. In the latter case, no permutation is needed if $B < A$, whereas a $2\leftrightarrow 3$ permutation is needed when $B > A$. In such a case, the permutation is not defined by the mass pattern alone. 

It is possible and useful to choose the ordering of leptons (and of their irreps) to start with in such a way to minimise the permutations needed. 

\subsection{H}
\label{sec:H}

We have now brought the lepton mass matrices in diagonal form, with the leptons in standard ordering
\begin{equation}
m_E = (P_E V_{e^c})^T \, m_E^\text{diag,so} \, (P_E V^{\phantom{T}}_e) ,
\qquad
m_\nu = (P_\nu D V_\nu)^T \, m_\nu^\text{diag,so} \, (P^{\phantom{T}}_\nu D V^{\phantom{T}}_\nu) .
\label{eq:so}
\end{equation}
A final point has to be taken into account in order to write the most general form of the PMNS matrix: the latter is not uniquely defined. This is because of the ambiguities associated to the definition of the mass eigenstates. The role of the unitary matrices $H$ is to take into account such ambiguities. 

In the real world case in which all the lepton masses are non-degenerate, the ambiguity is only associated to   unphysical phases. It is well known, for example, that the most general form of the CKM matrix contains five unphysical phases associated to the possibility to redefine the phases of up and down quarks, without modifying the diagonal form of the mass matrices. In the approximate world described by the symmetric limit, on the other hand, the ambiguity can be non-trivial, owing to the possible presence of degenerate, possibly vanishing, masses. It is then important to take into account such contributions, as they become physical when symmetry breaking effects, removing the degeneracy, are considered. 

The ambiguity affecting the definition of the PMNS matrix is associated to the unitary transformations $H_\nu$, $H_e$, $H_{e^c}$ leaving the diagonal form of the lepton mass matrices invariant, i.e.\ such that 
\begin{equation}
m_E^\text{diag,so} = H_{e^c}^T m_E^\text{diag,so} H^{\phantom{T}}_e, 
\qquad
m_\nu^\text{diag,so} = H^T_\nu m_\nu^\text{diag,so} H^{\phantom{T}}_\nu .
\label{eq:H}
\end{equation}
As only $H_e$ (and not $H_{e^c}$) enters the PMNS matrix, we are interested in the most general form of $H_e$ for which a proper $H_{e^c}$ exists satisfying \eq{H}. This taken into account, $H_e$ and $H_\nu$ are  characterised by
\begin{equation}
H_e (m_E^\text{diag,so})^2 = (m_E^\text{diag,so})^2 H_e, 
\qquad
m_\nu^\text{diag,so} = H^T_\nu m_\nu^\text{diag,so} H^{\phantom{T}}_\nu .
\label{eq:H2}
\end{equation}
In the previous equation, the eigenvalues in $m_E^\text{diag,so}$, $m_\nu^\text{diag,so}$ are supposed to be non-generic. We remind that our analysis focusses on a given mass pattern in table~\ref{tab:masses}, and that a set of eigenvalues in a certain pattern is generic if all the entries that are allowed to be different and non-zero are indeed different and non-zero. The possible forms of $H_e$, $H_\nu$ then only depend on the mass pattern being considered. Consider for example a mass pattern in which the mass eigenvalues are in the form in \eq{defpattern}, where the degeneracies are $d^E_0\ldots d^E_{N_E}$ for the charged leptons and  $d^\nu_0\ldots d^\nu_{N_\nu}$ for the neutrinos (the vanishing entries do not necessarily need to appear first, but let us for simplicity assume that this is the case). Then $H_e$ and $H_\nu$ have the form
\begin{equation}
H_e  = \bdiag(U_0, U_1, \ldots U_{N_E})  \;,
\qquad
H_\nu =  \bdiag(U'_0, R_1, \ldots R_{N_\nu}) \; ,
\label{eq:H3}
\end{equation}
where $U_i\in U(d^E_i)$, $U'_0\in U(d^\nu_0)$ are unitary matrices and $R_i \in O(d^\nu_i)$ are real orthogonal matrices. In \eq{H3}, $\bdiag$ denotes a block diagonal matrix, with the diagonal blocks specified as arguments. 

The $H_e$, $H_\nu$ contributions to the PMNS matrix have a different physical nature than the previous ones. The previous contributions are known, once the entries of the mass matrices in the symmetric limit are known. Barring special correlations, they correspond to large mixing if all the non-vanishing entries in the symmetric mass matrices are of the same order. On the contrary, $H_e$ and $H_\nu$ are unphysical, and undetermined, in the symmetric limit. However, they become physical (up to diagonal phases) after symmetry breaking effects split the degenerate mass eigenstates. By taking $H_e$ and $H_\nu$ into account, we then make sure that the PMNS matrix after  symmetry breaking is close to the one described by \eq{mostgeneral} in the symmetric limit, for some values of $H_e$, $H_\nu$. Depending on the specific form of the symmetry breaking effects, $H_e$ and $H_\nu$ can end up being be large, small, or zero. 

\subsection{The PMNS matrix}
\label{sec:combination}

By combining everything above, we find that the PMNS matrix is in the form in  \eq{mostgeneral}. That equation may contain some redundancy. The form of $V$ may have an undetermined component that can be parameterised by $H_e$ or $H_\nu$. This happens for example when $V$ is in principle non-trivial because of the presence of multiple copies of the same irrep, but those irreps correspond to massless leptons. We then choose $V$ to be the identity on the massless leptons and encode the undetermined component in $H_e$, $H_\nu$. Another redundancy appear in the case of Dirac structures, in which the diagonal neutrino mass matrix ends up having two degenerate eigenvalues. By definition, $H_\nu$ then contains a $2\times 2$ orthogonal rotation. However, as discussed in appendix~\ref{sec:D}, that rotation can be reabsorbed in a phase redefinition of $V$. We will therefore not include it in $H_\nu$. 

\bibliographystyle{JHEP}
\bibliography{refs}

\end{document}